\RequirePackage{fix-cm}
\documentclass[smallextended]{svjour3}       % onecolumn (second format)
\smartqed  % flush right qed marks, e.g. at end of proof
\usepackage[T1]{fontenc}
\usepackage[utf8]{inputenc}
\usepackage[english]{babel}
\usepackage{amssymb}
\usepackage{amsmath}
\usepackage{latexsym}
\usepackage{mathtools}
\usepackage{graphicx}
\usepackage{bm}
\usepackage{amsbsy}
\usepackage{wasysym}
\usepackage{booktabs}
\usepackage{caption}
\usepackage{tabularx}
\usepackage{epsfig}
\usepackage{natbib}
\usepackage{hyperref}
\usepackage[pagewise]{lineno}

\journalname{Experimental Astronomy}

\DeclarePairedDelimiter{\abs}{\lvert}{\rvert}

\begin{document}

\title{Orbit determination methods for interplanetary missions: development and use of the Orbit14 software}

\titlerunning{Orbit determination methods for interplanetary missions: the Orbit14 software}

\author{Giacomo Lari$^1$$^*$         \and
        Giulia Schettino$^1$$^2$     \and
        Daniele Serra$^1$        \and
        Giacomo Tommei$^1$
}


\institute{$^1$
              Dipartimento di Matematica, Università di Pisa, Largo Bruno Pontecorvo 5, Pisa, 56127, Italy\\
           $^2$
              IFAC-CNR, Via Madonna del Piano 10, Sesto Fiorentino (FI), 50019, Italy\\
           $^*$ \email{lari@mail.dm.unipi.it}  
}

\date{Received: date / Accepted: date}

\maketitle

\begin{abstract}
In the last years, a new generation of interplanetary space missions have been designed for the exploration of the solar system. At the same time, radio-science instrumentation has reached an unprecedented level of accuracy, leading to a significant improvement of our knowledge of celestial bodies. Along with this hardware upgrade, software products for interplanetary missions have been greatly refined. In this context,  we introduce Orbit14, a precise orbit determination software developed at the University of Pisa for processing the radio-science data of the BepiColombo and Juno missions. Along the years, many tools have been implemented into the software and Orbit14 capitalized the experience coming from simulations and treatment of real data.\\
In this paper, we present a review of orbit determination methods developed at the University of Pisa for radio-science experiments of interplanetary missions. We describe the basic theory of the process of parameters estimation and refined methods necessary to have full control on experiments involving spacecraft orbiting millions of kilometers far from the Earth. Our aim is to give both an extensive description of the treatment of radio-science experiments and step-to-step instructions for those who are approaching the field of orbit determination in the context of space missions. We show also the work conducted on the Juno and BepiColombo missions by means of the Orbit14 software. In particular, we summarize the recent results obtained with the gravity experiment of Juno and the simulations performed so far for the gravimetry-rotation and relativity experiments of BepiColombo.
\end{abstract}

\section{Introduction}
\label{sec:intro}
Since the 19th century, orbit determination (OD) allowed to reconstruct the orbit of celestial bodies starting from optical observations. They provided also first estimations of bodies' masses and shapes, even though affected by the limited accuracy of the telescopes of the time. With the starting of space exploration, scientists had the opportunity to collect observations exploiting the proximity of artificial objects to target bodies, like the Moon and planets. In particular, radio communications between ground-based stations and spacecraft provided high-precision Doppler data, which permitted to investigate the atmosphere and the gravitational field of celestial bodies. In this way, radio-science experiments broke through the barrier made of millions of kilometers of void space between the Earth and the other planets of the solar system. Prominent examples are the Mariner missions to the inner planets (Mercury, Venus and Mars) and the Pioneer and Voyager missions to the outer planets (Jupiter, Saturn, Uranus and Neptune). Through fly-bys of these bodies, scientists managed to greatly improve the knowledge of their masses and estimate their main gravitational anomalies (see, e.g., \citealt{CAMPBELL-SYNNOTT_1985,ANDERSON-etal_1987,CAMPBELL-ANDERSON_1989}). More recent missions provided spacecraft orbiting around planets, like MESSENGER to Mercury \citep{SANTO-etal_2001}, Mars Express to Mars \citep{CHICARRO-etal_2004}, Galileo to Jupiter \citep{JOHNSON-etal_1992}, Cassini-Huygens to Saturn \citep{MATSON-etal_2002} and many others. Orbiter missions around gas giants had the great opportunity to perform fly-bys of their moons, providing a precise estimation of the mass and gravitational field of Saturn's satellites \citep{JACOBSON-etal_2006} and the Galilean satellites \citep{SCHUBERT-etal_2004}, also through recent reanalyses of the data (see, e.g, \citealt{GOMEZCASAJUS-etal_2020}).

In the last years, OD reached unprecedented levels of accuracy, thanks to sophisticated radio-science instrumentation mounted on a new generation of spacecraft and extended ground-based station networks around the globe. In particular, a Ka-band Transponder (KaT) ensures a Ka/Ka link between the tracking station and the probe that allows a state-of-the-art accuracy in the Doppler data of $3\times 10^{-4}$ Hz over an integration time of $1000$ s. The advantage of this technology is the stability of the high-frequency signal ($32$-$34$ GHz), while the coupling with the X-band links allows to partially remove the noise due to the interaction with the space plasma \citep{ASMAR-etal_2005}. The KaT used on the Cassini-Huygens space mission allowed to estimate relativistic effects with an unprecedented accuracy \citep{BERTOTTI-etal_2003}. KaT technology delivered by Thales Alenia Space is mounted on the NASA Juno mission currently orbiting around Jupiter and on the ESA-JAXA BepiColombo mission currently flying toward Mercury \citep{CIARCIA-etal_2013}.

Precise observations are decisive to estimate with a great accuracy important physical and dynamical parameters of celestial bodies, and unveil tiny dynamical effects (see, e.g., \citealt{IESS-BOSCAGLI_2001}). Such parameters are essential for constraining interior models of planets (e.g., \citealt{HELLED_2018}) and minor bodies (e.g., \citealt{PATZOLD-etal_2016}), and for improving their orbits and generating new ephemerides (e.g., \citealt{VERMA-etal_2014}). However, this level of precision requires very accurate dynamical and observation models. In order to process radio-science data, space agencies and research institutes put great effort into the realization of sophisticated OD software. Examples are MONTE \citep{EVANS-etal_2018}, developed by JPL (Jet Propulsion Laboratory), and GINS \citep{PEROSANZ-etal_2011}, developed at CNES (Centre National d'\'Etudes Spatiales).

Since 2007, the Celestial Mechanics Group (CMG) of the University of Pisa, along with the spin-off SpaceDyS\footnote{www.spacedys.com}, developed a new precise OD code for interplanetary missions, called Orbit14. This project was born thanks to the financial support of the Italian Space Agency (ASI), which is deeply involved in the radio-science experiment of the BepiColombo space mission \citep{BENKHOFF-etal_2010}. Initially, the Orbit14 software was intended exclusively for BepiColombo. However, from 2011, because of the participation of ASI in the Juno space mission \citep{BOLTON-etal_2017}, the code was expanded in order to be used also for the Juno gravity experiment. In this way, not only does ASI provide advanced radio-science instrumentation to top space missions, but with Orbit14 it also ensures the capability of the data analysis.

Orbit14 is written in Fortran90 and its source code is fully accessible to our research group, so that it can be directly modified and updated. So, differently from other OD programs, where generally users have a software license, but they cannot easily modify the source code, with Orbit14 we have full synergy between the code development and its utilization. This provides a certain advantage in the experiments, as we can easily test new effects on the dynamics and observations. On the other hand, however, some effort is required to increase the versatility of the software, and its employment for setups very different from the nominal ones is not straightforward.

The CMG of Pisa worked on the development and implementation of Orbit14, providing also new dedicated methods for interplanetary missions. The code is able to deal with all the aspects of OD experiments: handling of input data, propagation of the dynamics, computation of the observables and estimation process. Moreover, several studies and simulations of the radio-science experiments of BepiColombo and Juno have been performed during the years. Starting from theoretic papers concerning the setup of the BepiColombo experiments \citep{MILANI-etal_2002}, high-quality formulation of computed data \citep{TOMMEI-etal_2010} and innovative multi-arc strategy \citep{ALESSI-etal_2012}, more recently, the CMG of Pisa has been able to perform full simulations of the BepiColombo gravity \citep{CICALO-etal_2016}, relativity \citep{SCHETTINO-etal_2018} and superior-conjunction \citep{SERRA-etal_2018} experiments, and Juno gravity experiment \citep{SERRA-etal_2016}. Orbit14 was used also for preliminary studies of the ESA JUICE mission \citep{GRASSET-etal_2013}, with a focus on spacecraft's orbit and the dynamics of the Galilean satellites \citep{DIRKX-etal_2017,LARI-MILANI_2019}. With the arrival of the Juno spacecraft at Jupiter and the collection of real data, Orbit14 supported the gravity experiment, contributing to the estimation of the asymmetric gravitational field of Jupiter \citep{IESS-etal_2018}. A detailed analysis of the first two gravity-dedicated perijoves (PJs) of Juno performed with Orbit14 was presented in \citet{SERRA-etal_2019}. The CMG of Pisa is carrying out the data processing along with the other research groups involved in the experiment, updating the gravitational parameters as Juno mission continues to orbit Jupiter \citep{DURANTE-etal_2020}.

This paper aims to provide a detailed description of the OD methods that have been implemented in the Orbit14 software in order to perform radio-scence experiments. We present also a review of the recent results obtained by the CMG of Pisa and we provide new insights about the last simulations and experiments performed with Orbit14. The article is organized as follows: in Sect.~\ref{sec:od}, we describe the basic theory of OD, while, in Sect.~\ref{sec:adv}, we expand it, reporting common and more recent techniques dedicated to radio-science experiments of interplanetary missions. In Sect.~\ref{sec:intermis}, we present detailed mathematical models for the dynamics of a spacecraft in the proximity of a celestial body and for the mathematical formulation of radio-science data. In Sect.~\ref{sec:juno}, we show the results of OD experiments performed with the Juno radio-science data, while, in Sect.~\ref{sec:bc}, we present the current state of the simulations of BepiColombo. Finally, in Sect.~\ref{sec:concl}, we summarize the methods and experiments described in the article and we present possible future perspectives for Orbit14.

\section{Orbit determination}
\label{sec:od}
Although we will focus on OD for radio-science experiments of interplanetary missions, in this section we want to give a generic formulation of the OD problem and its elements, as presented in \citet{MILANI-GRONCHI_2010}.

\begin{figure}[t]
\includegraphics[scale=0.4]{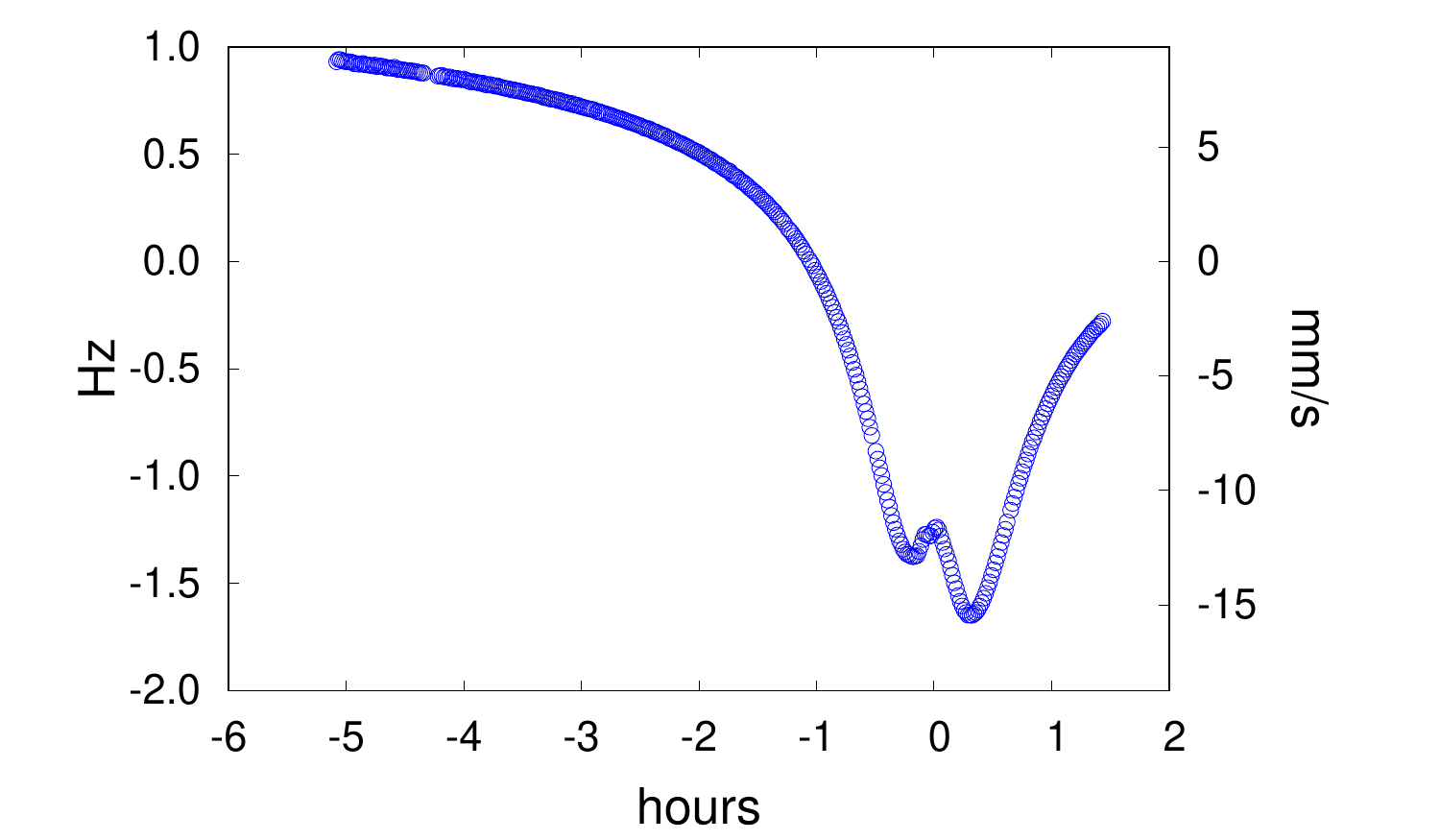}
\hspace{-0.5cm}\includegraphics[scale=0.4]{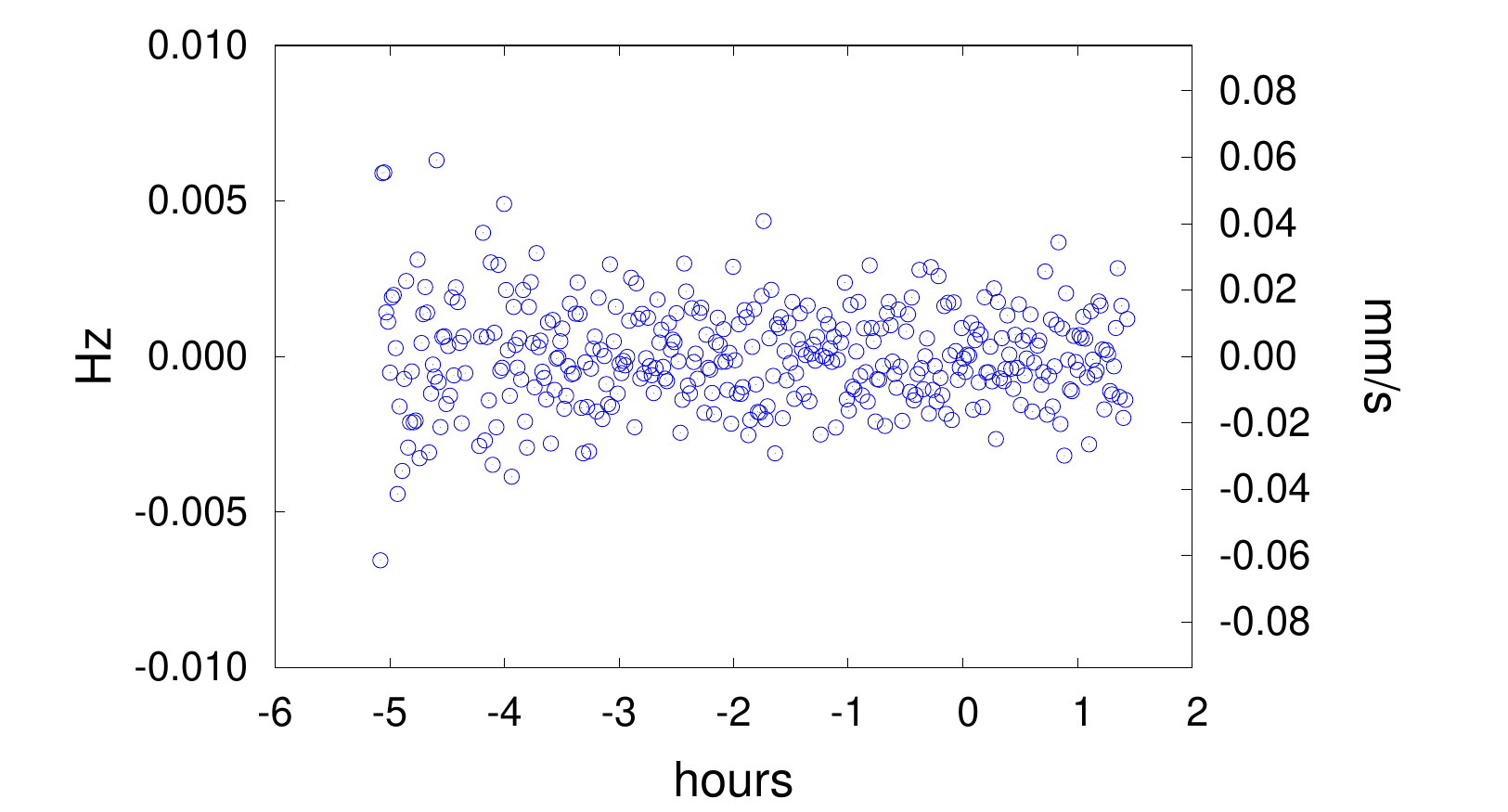}
\caption{Pre-fit (left) and post-fit (right) residuals of the Juno Doppler data during PJ03. In the pre-fit residuals, the main source of discrepancy between the observed and computed data is the initial state of the spacecraft, which is taken from SPICE kernels provided by the mission and then corrected in the orbit determination process.}
\label{fig:residuals}
\end{figure}

The aim of OD is to estimate some parameters that define or influence the motion of a celestial body, starting from a suitable set of observations. In this context, the determination is not limited to the orbital elements (or, equivalently, the position and velocity) of the observed body, but possibly includes physical parameters of a target body, such as its mass.

We consider $m$ observations $r_i$ of a certain celestial or artificial body taken at times $t_i\ (i=1,m)$. In principle, the observations $r_i$ can be of very different nature. For instance, they can be angular measurements collected by a telescope or distances obtained with a radar. For radio-science experiments, we are mainly interested in the Doppler, which is the variation in frequency of the radio signal because of the non-null velocity of the emitter (e.g., a ground-based station) and the receiver (e.g., the spacecraft). We refer to all these quantities as observed data or observed observables. We can define a mathematical model $R$ in order to predict the numerical outcome of the observation $r_i$ at time $t_i$
\begin{equation}
\label{eq:predict}
r(t)=R(t,\mathbf y(t),\boldsymbol{\nu}),
\end{equation}
which depends on time $t$, state vector $\mathbf y$ (positions and velocities) of the bodies involved in the observation and a certain number of parameters $\boldsymbol{\nu}$ that affect the measurement, called kinematic parameters. The evaluation of Eq.~\eqref{eq:predict} at the time $t_i$ of the $i$th observation provides the quantity $r(t_i)$, called computed data or computed observable, that aims to reproduce the actual value obtained with the measurement.

The vector $\mathbf y(t)$ depends on time, as the state of the bodies changes with the motion. Therefore, in order to compute $r(t)$, we need a dynamical model for propagating the involved bodies:
\begin{equation}
\label{eq:dyneq}
\begin{cases}
\dot{\mathbf {y}}=\mathbf f(t,\mathbf {y},\boldsymbol{\mu}),\\
\mathbf {y}(t_0)=\mathbf {y}_0.
\end{cases}
\end{equation}
Eq.~\eqref{eq:dyneq} defines a Cauchy problem with initial conditions $\mathbf {y}_0$. The function $\mathbf f$ depends on time, on the states of the bodies and on other parameters $\boldsymbol{\mu}$ that influence the motion, called dynamical parameters. Once we have integrated the system in Eq.~\eqref{eq:dyneq}, we are able to calculate $r(t_i)$ through Eq.~\eqref{eq:predict}.

From the observed and computed data, we obtain the vector of the residuals 
\begin{equation}
\label{eq:residuals}
\boldsymbol \xi=(\xi_i)_{(i=1,m)},\qquad \text{where}\quad \xi_i=r_i-r(t_i),
\end{equation}
which describes how our predictions differ from the actual measurements. In principle, the residuals depend on all the parameters we have introduced, i.e., $\mathbf y_0,\ \boldsymbol{\mu},\ \boldsymbol{\nu}$. However, we can choose a limited subset of parameters that we want to actually determine with our experiment, while we consider the others as sufficiently known and we keep them fixed. We introduce the vector $\mathbf x$ of the fit parameters, or solve-for parameters, and we can write $\boldsymbol \xi=\boldsymbol \xi(\mathbf x)$.

In order to estimate the fit parameters, we define a target function $Q(\boldsymbol \xi(\mathbf x))$ in such a way that it is convex with respect to the residuals. Our goal is to minimize $Q$ (hence the residuals) and to find the corresponding minimum point $\mathbf x^*$. This procedure is called least squares method: the expression we use for the target function is
\begin{equation}
\label{eq:target}
Q=\frac{1}{m}\boldsymbol{\xi}^TW\boldsymbol{\xi},
\end{equation}
where $W$ is the $m\times m$ matrix of the weights of the observations. It describes the fact that data are not uniform. Moreover, if $W$ is not diagonal, it accounts for correlations between different observations. For the rest of this article, we will consider $W=\mathrm{diag}(w_i)$, with $w_i=1/\sigma_i^2$ and $\sigma_i$ the nominal accuracy of the $i$th observation. We note that $W$ can simply account for a change of units, in order to handle different sets of data, or it allows to discriminate the observations with respect to their level of accuracy. This is very common in astrometry, where observations can be decades apart, but also in radio-science, where the weather over the tracking station can affect the measurements differently from one day to another.

In order to find the minimum point $\mathbf x^*$, we derive Eq.~\eqref{eq:target} and we impose that it is equal to $\mathbf 0$:
\begin{equation}
\label{eq:derivtarget}
\frac{\partial Q(\mathbf x^*)}{\partial \mathbf x}=\mathbf 0\quad\to\quad\frac{2}{m}B^TW\boldsymbol{\xi}=\mathbf 0.
\end{equation}
The $m\times N$ matrix $B=\partial \boldsymbol{\xi}/\partial \bf x$ is called design matrix and it describes how the residuals vary when the fit parameters change in value. From the definition of residuals given in Eq.~\eqref{eq:residuals}, if $p$ is one of the fit parameters, we have that
\begin{equation}
\label{eq:derres}
\frac{\partial \boldsymbol{\xi}}{\partial p} = -\frac{\partial {R}(t,\mathbf y,\boldsymbol{\nu})}{\partial p}.
\end{equation}
In case $p$ is a kinematic parameter, the derivative is straightforward. Instead, if $p$ is included in $\boldsymbol \mu$ or $\mathbf y_0$, Eq.~\eqref{eq:derres} can be expanded as
\begin{equation}
\frac{\partial \boldsymbol{\xi}}{\partial p} = -\frac{\partial {R}}{\partial \mathbf y(t)}\frac{\mathbf y(t)}{\partial p},
\end{equation}
because $\mathbf y(t)$ depends on dynamical parameters and initial conditions as described in Eq.~\eqref{eq:dyneq}. Therefore, in the OD process, not only do we propagate the state of the bodies through Eq.~\eqref{eq:dyneq}, but it is necessary to propagate the differential equations of the partial derivatives $\partial \mathbf{y}/\partial p$ through the system
\begin{equation}
\label{eq:parder}
\begin{cases}
\displaystyle\frac{d}{dt}\frac{\partial \mathbf {y}}{\partial p}=\frac{\partial \mathbf f}{\partial \mathbf {y}}\frac{\partial \mathbf {y}}{\partial p}+\frac{\partial \mathbf f}{\partial p},\\
\displaystyle\frac{\partial \mathbf {y_0}}{\partial p}=\mathbf a.
\end{cases}
\end{equation}
If $p$ belongs to $\boldsymbol \mu$, then $\mathbf a=\mathbf 0$; while, if $p$ belongs to $\mathbf y_0$, $\mathbf a$ has 0 in each component apart from a $1$ in correspondence of the index of $p$ in $\mathbf y_0$.

In order to solve the system in Eq.~\eqref{eq:derivtarget}, we can use the Newton method. At iteration $k+1$, we have
\begin{equation}
\label{eq:expdertarget}
\mathbf 0=\frac{\partial Q(\mathbf x^*)}{\partial \mathbf x}=\frac{\partial Q(\mathbf x_k)}{\partial \mathbf x}+\frac{\partial^2 Q(\mathbf x_k)}{\partial \mathbf x^2}(\mathbf x^*-\mathbf x_k)+\mathcal O (\abs{\mathbf x^*-\mathbf x_k}^2),
\end{equation}
where
\begin{equation}
\label{eq:approxnormal}
\frac{\partial^2 Q(\mathbf x_k)}{\partial \mathbf x^2}=\frac{2}{m}(B^TWB+\boldsymbol \xi^T WH)\approx\frac{2}{m}B^TWB
\end{equation}
and $H=\partial^2 \boldsymbol{\xi}/\partial \bf x^2$. The approximation in Eq.~\eqref{eq:approxnormal} holds if the residuals are small, which is the case if the computed data are not so distant from the observed ones.

\begin{figure}[t]
\centering
\includegraphics[scale=0.45]{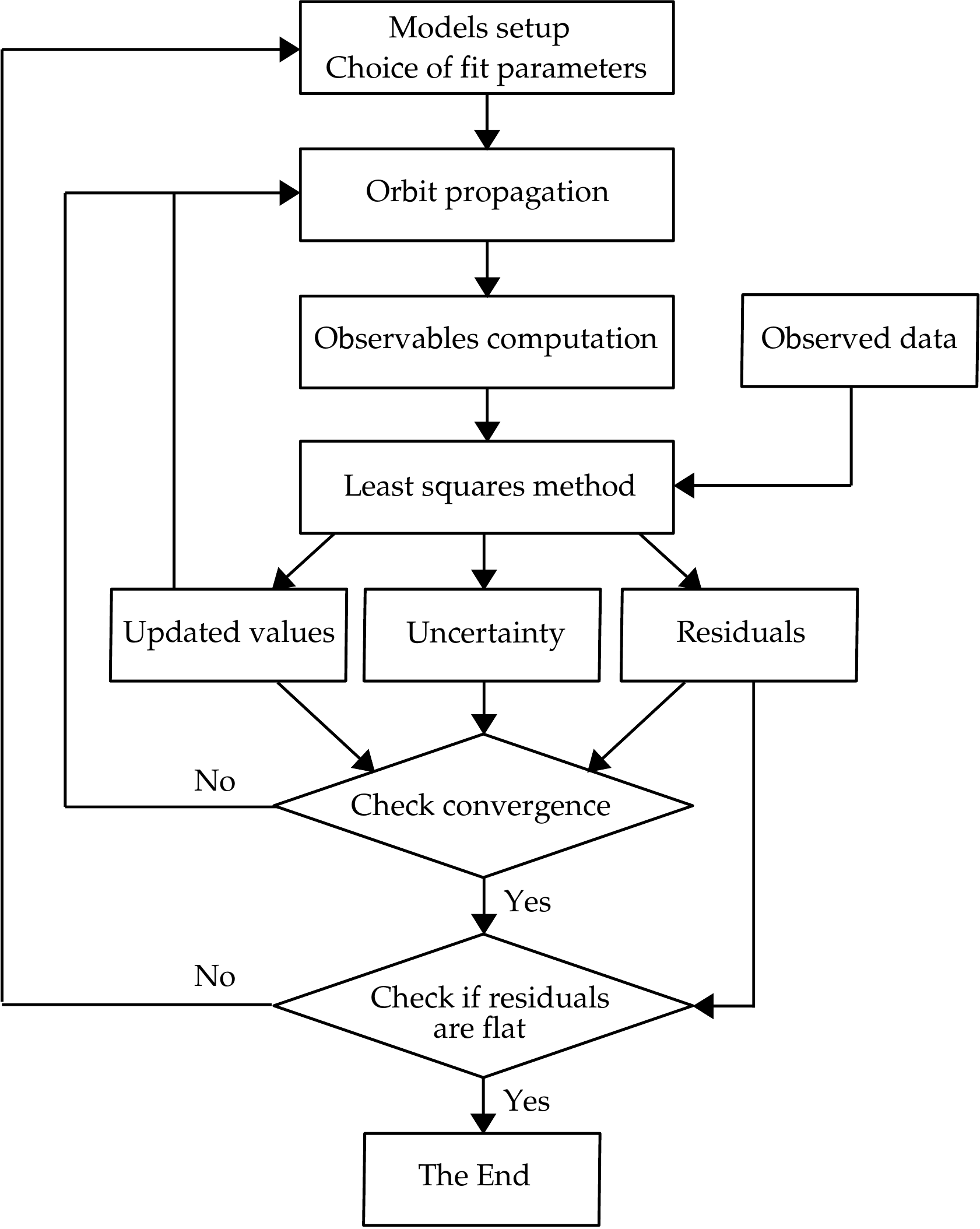}
\caption{Block diagram of orbit determination experiments.}
\label{fig:block}
\end{figure}

Using the approximations described in Eqs.~\eqref{eq:expdertarget} and~\eqref{eq:approxnormal}, we can compute the updated value $\mathbf x_{k+1}$ of the fit parameters solving the system of linear equations
\begin{equation}
\label{diffcorr}
C(\mathbf x_{k+1}-\mathbf x_{k})=D.
\end{equation}
Such variant of the Newton method is called differential correction. The positive semi-definite $N\times N$ matrix $C=B^TWB$ is called normal matrix, while $D=-B^T\boldsymbol \xi$ is the constant term of the equation. Both $C$ and $D$ depend on $\mathbf x_k$. At convergence of this iterative procedure, we obtain the final value of the fit parameters. If we start from a good first guess $\mathbf x_0$ and the matrix $C$ is non-degenerate, the last value of $\mathbf x_{k+1}$ should be very close to the real minimum $\mathbf x^*$ and the residuals should be small, and hopefully flat. An example of residuals before and after the differential corrections is shown in \figurename~\ref{fig:residuals}.

\begin{table}[t]
{\small
\begin{center}
\noindent\begin{tabular}{lccl}
\toprule
iter. & $Q$ & $\| \mathbf x_{k+1}-\mathbf x_{k} \|_C$ & convergence\\
\midrule
$1$  & $2.772\times 10^5$ & $1.273\times{10^3}\phantom{^-}$ & no\\
$2$  & $1.050 \phantom{\times 10^{000}}$ & $2.378\times{10^{-1}}$ & no\\
$3$  & $1.041 \phantom{\times 10^{000}}$ & $1.509\times{10^{-2}}$ & yes ($<5\times 10^{-2}$) \\
\bottomrule
  \end{tabular}
\end{center}
}
\caption{Target function $Q$ and norm of the parameters' variation for the different iterations of an orbit determination experiment performed with the PJ03 Doppler data of Juno, whose accuracy is set to $\sigma_i\approx 1.6\times 10^{-3}$ Hz.}
\label{tab:conv}
\end{table}

As the target function in Eq.~\eqref{eq:target} is normalized with respect to the number of the observations and to their accuracy, at convergence we expect a value of $Q$ close to $1$. This is possible if our models manage to absorb all the systematic signals, dynamical and kinematic, in the residuals. We cannot hope to reach $Q\approx 0$, as we cannot obviously absorb the white noise in the residuals due to the non-perfect accuracy of the measurements (see post-fit residuals in \figurename~\ref{fig:residuals}).

The convergence criteria of the iterative process in Eq.~\eqref{diffcorr} are based on the variation of the values of $\mathbf x$ and $Q$. For the fit parameters, we need a suitable norm to discern if $\mathbf x_{k+1}-\mathbf x_{k}$ is small or not. Following \citet{MILANI-GRONCHI_2010}, we define
\begin{equation}
\| \mathbf b \|_C=\sqrt{\frac{\mathbf b^T C \mathbf b}{N}}.
\end{equation}
We declare convergence at iteration $k+1$ if
\begin{itemize}
\item $\| \mathbf x_{k+1}-\mathbf x_{k} \|_C \ll 1$, or
\item $Q_{k+1}-Q_k \ll 1$ for a certain number of iterations.
\end{itemize}
In the first case, the small variation in the value of the fit parameters means that we are already close to the stationary point and other iterations do not bring further significant improvement to the solution. Instead, the second criterion allows to stop the differential corrections in cases where the sequence $\{\mathbf x_k\}_k$ is near the stationary point but it keeps moving around it, because of the flatness of the function $Q$ along some directions around $\mathbf x^*$, given for example by strong correlations between fit parameters. 

In \tablename~\ref{tab:conv}, we reported the value of the target function and the norm of the parameters' variation obtained fitting Juno Doppler data collected at PJ03. Convergence occurs at iteration $3$ because of the small variation in the fit parameters' value (threshold set to $5\times 10^{-2}$).

\begin{figure}[t]
\centering
\includegraphics[scale=0.7]{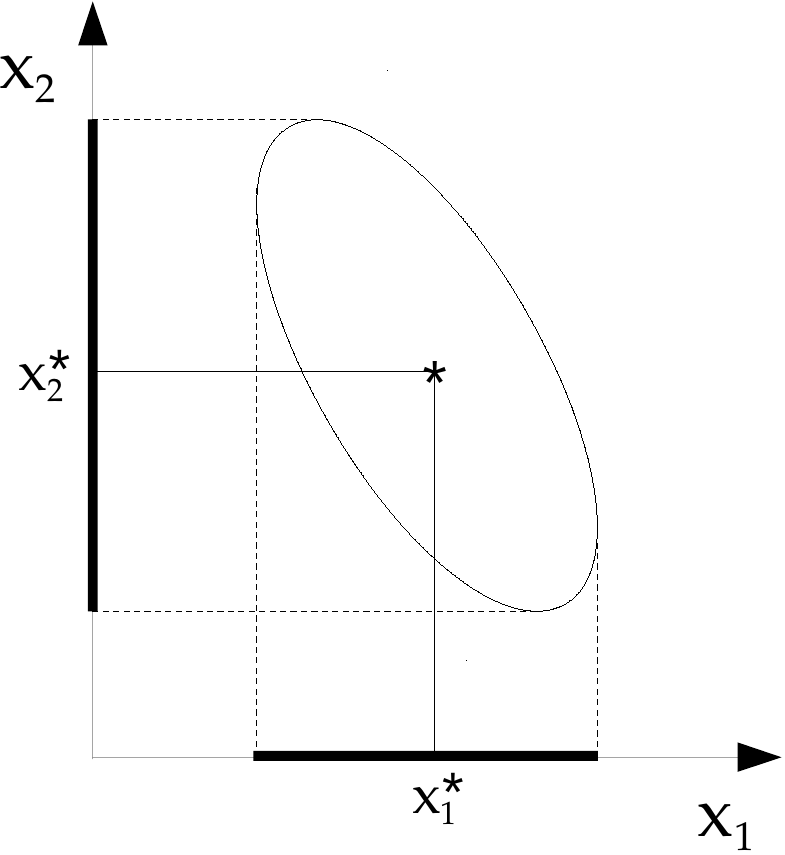}
\caption{Example of a confidence ellipsoid for a two-dimensional parameter space. The ellipsoid has an axis larger than the other, and its inclination is caused by a non-zero correlation between the parameters. In the center, we have the nominal solution $\mathbf x^*$, while the ellipsoid projections on the Cartesian axes represent the formal uncertainties of the parameters.}
\label{fig:confell}
\end{figure}

Apart from computing new values for the fit parameters, the OD process provides also a formal uncertainty of the found solution. Indeed, in a neighborhood of the nominal solution $\mathbf x^*$, the target function can be expressed as
\begin{equation}
 Q(\mathbf x)= Q(\mathbf x^*)+\Delta  Q,
\end{equation}
where $\Delta  Q$ is called penalty. From Eqs.~\eqref{eq:expdertarget} and~\eqref{eq:approxnormal}, we have
\begin{equation}
\Delta  Q \approx (\mathbf x^*-\mathbf x)^T C (\mathbf x^*-\mathbf x).
\end{equation}
We can define a region of the parameters' space for which the penalty does not exceed a certain control threshold $\sigma$:
\begin{equation}
\label{eq:confidence}
Z(\sigma)=\{\mathbf x\ |\ (\mathbf x^*-\mathbf x)^T C (\mathbf x^*-\mathbf x)\le \sigma^2\}.
\end{equation}
As $C$ is positive definite (in a non-degenerate case), $Z(\sigma)$ describes an ellipsoid (see the example in \figurename~\ref{fig:confell}). When we fix a value of $\sigma$, we consider acceptable all the values of the fit parameters inside $Z(\sigma)$. A large $\sigma$ defines a wide confidence region, with a consequent large formal uncertainty in the determination of the parameters, while a small $\sigma$ defines a strict confidence region, which could lead to a too optimistic estimation. Moreover, $Z(\sigma)$ can have axes larger than others and an elongated shape toward their directions, meaning that the corresponding parameters are determined worse than the others.

This is formally described by $\Gamma=C^{-1}$, called covariance matrix. By definition of the confidence ellipsoid in Eq.~\eqref{eq:confidence} with $\sigma=1$ ($1$-$\sigma$ uncertainty), this matrix contains the uncertainty of the parameters and their correlations:
\begin{equation}
\label{eq:formunc}
\begin{aligned}
\sigma(x_i)&=\sqrt{\Gamma_{i,i}} \quad (i=1,N),\\
\mathrm{corr}(x_i,x_j)&=\Gamma_{i,j}/\sqrt{\Gamma_{i,i}\Gamma_{j,j}} \quad (i,j=1,N).
\end{aligned}
\end{equation}
If $\sigma\ne 1$, we can just set $\Gamma=\sigma^2 C^{-1}$ and use the same formulas. In general, it is assumed a normal distribution, so that $1$-sigma corresponds to a probability of around $68\%$ that the real solution lies in the confidence ellipsoid and $3$-sigma around $99.7\%$.

In the case of two parameters $x_1$ and $x_2$, we can easily draw their confidence ellipsoid as done in \figurename~\ref{fig:confell}. Given their covariance matrix
\begin{equation}
\Gamma=
\begin{bmatrix}
\Gamma_{11} & \Gamma_{12} \\
\Gamma_{21} & \Gamma_{22}
\end{bmatrix},
\end{equation}
which can be extracted from a bigger matrix if the number of parameters is larger than two, we compute its two positive eigenvalues $\lambda_1$ and $\lambda_2$, and the angle $\alpha$ between the $x$-axis and the eigenvector corresponding to the largest eigenvalue. The $1$-$\sigma$ confidence ellipsoid is the ellipse centered in the estimated values $\approx(x_1^*,x_2^*)$ of the parameters, with axes $2\sqrt{\lambda_1}$ and $2\sqrt{\lambda_2}$ long and rotated by an angle equal to $\alpha$.

The covariance matrix is essential as it describes the quality of the estimation of parameters. In the first stages of a mission, when real data are not yet available, the computation of the covariance matrix is extremely important to evaluate mission's goals and performances. The standard procedure is to simulate the data following the trajectory and design of the mission, and to compute the covariance matrix given by the simulated OD experiment. In this way, it is possible to preliminarily assess the level of accuracy with which the fit parameters will be estimated, considering all the possible scenarios.

\section{Advanced theory}
\label{sec:adv}
In Sect.~\ref{sec:od}, we presented the basic theory of OD, but that is not the whole story. For example, the algorithm we described can not work if some numerical problems occur in the computations. It is the case when the normal matrix $C$ is bad conditioned, so that we cannot invert it and solve the system in Eq.~\eqref{diffcorr}. In this section, we want to describe some tools that can help to deal with such obstacles, along with some common strategies for radio-science experiments.

\subsection{Apriori}
\label{subsec:adv_ap}
As already stated, the aim of OD experiments is to use a set of data to determine some parameters of interest. However, in many cases, we already have a preliminary estimate $\mathbf x^P$ of the values of the fit parameters $\mathbf x$. This knowledge can come from previous experiments performed with other sets of data (taken from Earth or past space missions) or from theoretical studies. Indeed, both sources are very useful not only for building the first guess $\mathbf x_0$ of the iterative process in Eq.~\eqref{diffcorr}, but also to constrain the new solution.
 
More precisely, this apriori information can be added to the OD process as $m_P$ further observations. We define the $N\times N$ apriori normal matrix
\begin{equation}
C^P=\mathrm{diag}((\sigma_i^P)^{-2}),
\end{equation}
where $\sigma_i^P$ $(i=1,m_P)$ is the uncertainty with which the parameters are considered already known. In the case we do not have (or we do not want to use) apriori information for some parameters, we can just set to $0$ the corresponding weight in $C^P$ ($\sigma_i^P=\infty$). Then we add to the target function a new piece that we require to minimize
\begin{equation}
 Q_P(\mathbf x)=\frac{1}{m_P}(\mathbf x-\mathbf x^P)^T C^P(\mathbf x-\mathbf x^P).
\end{equation}
The new normal matrix reads
\begin{equation}
\label{eq:apriorinormal}
C=C^P+B^TWB.
\end{equation}

The purpose of including apriori information is to add a constraint which prevents the new solution from moving away from predetermined values of the parameters. In the case we obtain a new formal uncertainty for a parameter which is largely below its apriori uncertainty, the apriori contribution is generally minimal. Instead, for parameters that do not gain a large improvement, or even are poorly determined by the experiment, apriori have a great impact on the solution. As a matter of fact, they can be an essential tool to make the normal matrix $C$ invertible and solve bad conditioning problems (see Sect.~\ref{subsec:adv_rk}). 

Apriori are also very useful to investigate tiny dynamical effects. If we suppose that the effect linearly depends on the unknown parameter $x_{\overline{k}}$, we can set $x_{\overline{k}}^P=0$ and assign a small apriori uncertainty. In this way, we introduce the effect associated to $x_{\overline{k}}$ into the fit, but we limit its impact. This strategy is adopted, for example, with the inclusion of random accelerations in the dynamical model of a spacecraft around a celestial body, in such a way to absorb small remaining signals in the Doppler residuals (see \citealt{IESS-etal_2019} and \citealt{DURANTE-etal_2020}).

\subsection{Cure to rank deficiencies}
\label{subsec:adv_rk}
Sometimes, not only is the normal matrix not invertible because of lack of observations, but because the experiment is not sensitive to some parameters. This can happen when a dynamical effect we are investigating is too weak. However, in some cases, the observations are not able to reveal the change of value of a certain parameter because of a purely geometric problem.

More precisely, there could exist a subspace $K$ of the parameters' space such that
\begin{equation}
\label{eq:rankdef}
B\mathbf v=\mathbf 0,\qquad \forall\  \mathbf v\in K.
\end{equation}
Therefore, by definition of the design matrix $B$, the change of the residuals due to a shift $s\mathbf v$ in the parameters results to be of the second order in $s$:
\begin{equation}
\boldsymbol \xi(\mathbf x+s\mathbf v)-\boldsymbol \xi(\mathbf x)=sB\mathbf v + \mathcal O(s^2)=\mathcal O(s^2).
\end{equation}
This means that the observations are not able to constrain the value of some parameters. Moreover, Eq.~\eqref{eq:rankdef} implies that the normal matrix $C$ is not invertible and has a rank deficiency of the order of the dimension of $K$. 

Rank deficiencies are related to geometric symmetries arising in the configuration of the problem we are studying (for details, see \citealt{MILANI-GRONCHI_2010}, Chapt. 6). Sometimes we notice the presence of a rank deficiency because we are unable to invert $C$, but the symmetry remains concealed. Let $G$ be a group of transformations of the fit parameters' space. We call $G$ a group of exact symmetries if
\begin{equation}
\boldsymbol \xi(g(\mathbf x))=\boldsymbol \xi(\mathbf x), \qquad \forall\ g\in G;
\end{equation}
i.e., the residuals are invariant for every action $g$ of the group $G$.

Examples of exact symmetries in the context of space missions are described in \citet{BONANNO-MILANI_2002} and \citet{MILANI-GRONCHI_2010}, Chapt. 6. Here, we present two symmetries which have been investigated for the BepiColombo mission (see \citealt{MILANI-etal_2002}).

We consider the ideal 4-body system composed by the Sun, Mercury, Earth and a spacecraft orbiting around Mercury. Once we fix the origin, the only parameters involved are the initial conditions of Mercury $\mathbf y_0^\text{M}$, of the Earth $\mathbf y_0^\text{E}$, of the spacecraft $\mathbf y_0$ and the gravitational parameters (or the masses) of the three celestial bodies. We assume that the only measurements of the probe are its distance from the Earth and the time variation of this distance (range and range-rate, see Sect.~\ref{sec:intermis}). If we try to estimate $\mathbf y_0^\text{M}$ and $\mathbf y_0^\text{E}$ (total dimension $12$) with the OD process presented in Sect.~\ref{sec:od}, we find a rank deficiency of order $3$ in the matrix $C$. This deficiency is due to a symmetry for rotations: if we take $G=SO(3)$, which is the group of the rotation of the three-dimensional space, it is easy to show that the residuals do not change for the action of every $g\in SO(3)$ on the space of the parameters. This results from the fact that both equations of the motion of point masses and the considered observations are invariant with respect to the rotation of the space.

Instead, if we assume to have only angular observations of the spacecraft, we find another exact symmetry, this time of dimension $1$, which links the gravitational parameter of the Sun and the distances of the bodies. It is the so-called symmetry for scaling and it is due to the fact that if we change all the lengths of the system by a factor $\lambda$, all the masses by a factor $\mu$ and all the times by a factor $\tau$, in such a way that they verify Kepler's third law $\lambda^3=\tau^2\mu$, then the equations of motion remain unchanged. When considering only angular observations, the change of scale remains undetected by the residuals.

The examples given above refer to simplified models of real configurations. In more complete models, exact symmetries are generally broken. For example, because of perturbations present in the dynamical system, the invariance of the equations of motion stops to be valid. However, as these effects are small, an approximate symmetry remains: the residuals under the action of the elements $g$ of the group are not invariant, but they have just a tiny change
\begin{equation}
\boldsymbol \xi(g(\mathbf x))\approx\boldsymbol \xi(\mathbf x)+\epsilon \mathbf a;
\end{equation}
with $\abs{\mathbf a}=1$ and $\epsilon\ll 1$ a small parameter associated to the perturbation.

Approximate symmetries imply that for some directions $\mathbf v$
\begin{equation}
\label{eq:rankdefapp}
B\mathbf v=\epsilon\mathbf a,
\end{equation}
so that matrix $C$ is not degenerate, but it is still bad conditioned. In this case, we say that we are in presence of an approximate rank deficiency.

Therefore, even if in realistic situations we do not find exact symmetries and eigenvalues equal to $0$ in the normal matrix, we still face a computational obstacle which needs to be overcome. Several possible solutions have been proposed and adopted during the years and here we give a summary, with a special focus on the radio-science experiment of BepiColombo (see Sect.~\ref{subsec:bc_rk}).

First, as symmetries depend on the observations, considering different kinds of data in the OD allows to break them. Although radio-science experiments collect radial measurements, angular data of the spacecraft can be provided by other experiments, such as VLBI (see \citealt{DUEV-etal_2012}). Obviously, this is not always possible; therefore, other solutions must intervene directly on the computation scheme. For example, we can remove some parameters involved in the symmetries from the list of the fit parameters (descoping); as their impact on the residuals is limited, this should not compromise the quality of the estimation. In \citet{MILANI-etal_2002}, descoping was adopted, removing $4$ coordinates of $(\mathbf y_0^\text{M},\mathbf y_0^\text{E})$ from the fit parameters. However, this is a real drastic choice, which can be softened: we can add some targeted apriori conditions on the parameters we know are problematic, as described in Sect.~\ref{subsec:adv_ap}. More sophisticated solutions provide constraints on the parameters which limit the action of the group of symmetry. For example, \citet{SCHETTINO-etal_2018} developed conditions to be added in the fit, in order to inhibit the rotations around the Cartesian axes. In Sect.~\ref{subsec:bc_rk}, we will discuss in more details this simmetry issue in the case of the BepiColombo radio-science experiment.

\subsection{Multi-arc strategy}
\label{subsec:adv_mult}
In the context of space missions, when we have sets of data separated in time (sometimes even by weeks or months), a useful approach for simplifying the computation is to divide the observations in $n$ groups. This is a very common strategy, as high-quality radio-science data are collected when the spacecraft is very near to its target body, such as at the pericenter of its orbit or during a fly-by, because they are more sensitive to the dynamical parameters. For each group, we consider its own initial conditions vector $\mathbf y_0^{k}$ of the spacecraft and we propagate it along the time span which encloses the observations of the group. Therefore, although it exists a whole orbit which covers the entire duration of the mission, we consider instead $n$ arcs independent from each others. The motivation of this strategy is that, generally, we are not interested in retrieving the orbit of the spacecraft when it is far from the target body, while we want to obtain a good estimation of the position and velocity of the probe just when the observations are collected.

Moreover, typically, it is almost impossible to compute a years-long global orbit, because of unmodeled dynamical effects occurring far from the observation periods (such as maneuvers or non-gravitational forces). Sometimes multi-arc strategy is unavoidable when dealing with radio-science experiments. It is the case of space missions with multiple fly-bys, like Cassini to Saturn and JUICE to Jupiter. As for the latter, \citet{LARI-MILANI_2019} showed that it is impossible to retrieve the whole orbit of the probe starting from a single set of initial conditions, as the close approaches with the moons of Jupiter rapidly introduce chaos in the motion, nullifying the effort to propagate accurately the whole tour of the mission once for all.

In the multi-arc approach, we can divide the vector of the fit parameters in two sub-vectors $\mathbf x=[\mathbf g, \mathbf l]$, where $\mathbf g$ are the global parameters and $\mathbf l$ are the local parameters. With the term global, we mean that they are parameters which affect and intervene in all the arcs, while, with local, we indicate parameters that are exclusive of their corresponding arc. Local parameters can be further subdivided according to the arcs $\mathbf l=[\mathbf l_i]_{(i=1,n)}$. Vector $\mathbf l_i$ includes the initial conditions of the spacecraft of the $i$th arc and eventually other parameters that affect only that arc.

Consequently, the residuals of a certain arc are independent from local parameters of another arc. Therefore, the normal matrix $C$ has a particular arrow shape, as the blocks
\begin{equation}
\label{eq:localind}
C_{\mathbf l_i\mathbf l_j}=\left(\frac{\partial \boldsymbol\xi}{\partial \mathbf l_i}\right)^TW\frac{\partial \boldsymbol\xi}{\partial \mathbf l_j}=\mathbf 0, \quad \text{if }i\ne j.
\end{equation}

This allows to follow a particular scheme (see \citealt{MILANI-GRONCHI_2010}, Chapt. 15) to solve the linear system defined in Eq.~\eqref{diffcorr}, which is computationally less expensive than the direct inversion of the normal matrix $C$. Such a scheme, that was convenient in the past, is not necessary anymore with the current computational power of the modern CPUs.

A weakness of the classic multi-arc is that it gets rid of the accumulated information contained in the global orbit of the spacecraft. A more sophisticated approach is the so-called constrained multi-arc strategy, developed by \citet{ALESSI-etal_2012}. In this case, local spacecraft's states are not propagated within the observations' time span only, but for a much longer period. Taken two subsequent arcs $j$ and $j+1$, we define the conjunction time $t_c^{j}$ as the time in the middle between the two arcs, and we propagate the spacecraft up to that point, both forward from arc $j$ and backward from arc $j+1$. As the new arcs are longer, they are called extended arcs.

\begin{figure}[t]
\centering
\includegraphics[scale=0.5]{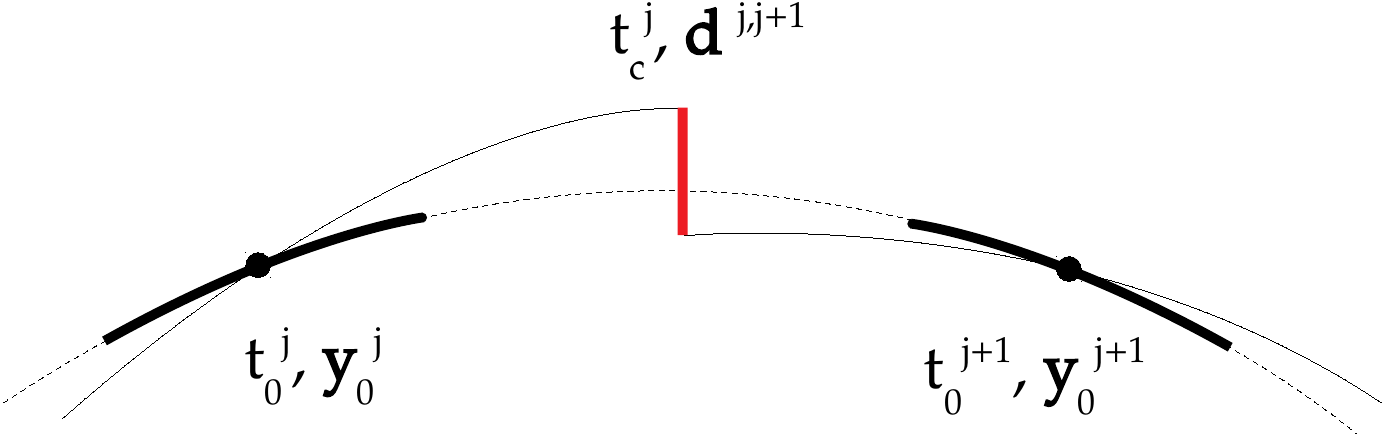}
\caption{Schematic representation of a jump between two subsequent extended arcs. The goal of the constrained multi-arc strategy is to recover a spacecraft's orbit closer to the real one through the minimization of the jumps.}
\label{fig:jump}
\end{figure}

We define the jump between two extended arcs as the discrepancy between the two computed states of the probe at the conjunction time
\begin{equation}
\label{jump}
\mathbf d^{j,j+1}=\Phi^{t_c^{j}}_{t_0^{j+1}}(\mathbf y_0^{j+1})-\Phi^{t_c^j}_{t_0^j}(\mathbf y_0^j),
\end{equation}
where $\Phi^{t}_{t_0}(\mathbf y_0)$ is the solution of Eq.~\eqref{eq:dyneq} (relative to the spacecraft) at time $t$ starting from $t_0$ and initial condition $\mathbf y_0$ (see \figurename~\ref{fig:jump}). The vector $\mathbf d^{j,j+1}=\mathbf d^{j,j+1}(\mathbf x)$ depends on the global dynamical parameters, but also on the local parameters of arcs $j$ and $j+1$. In the constrained multi-arc approach, we want to minimize the jumps, in order to recover the information contained in the non-observed part of spacecraft's orbit.

Therefore, we add a new contribution to the target function: 
\begin{equation}
\label{qjump}
 Q_d=\frac{1}{m_d}\frac{1}{\mu}\sum_{j=1}^{n-1}\left(\mathbf d^{j,j+1}\right)^T C^{j,j+1}\mathbf d^{j,j+1},
\end{equation}
which must be taken into account in the computation of the normal matrix, as the new design matrix $B=\partial(\boldsymbol \xi,\mathbf d)/\partial \mathbf x$ includes also the derivatives of the jumps. In Eq.~\eqref{qjump}, matrix $C^{j,j+1}$ is a weight matrix that sets the desired level of the jumps' size (e.g., meters in position, centimeters per second in velocity), while $\mu$ is a penalty parameter which fixes the global weight of the jumps constraint in the target function. The integer $m_d$ is the total number of constraints, i.e. $6(n-1)$ if we want to minimize all the six-dimensional jumps between the arcs.

From Eqs.~\eqref{jump} and~\eqref{qjump}, we have that $C_{\mathbf l_j\mathbf l_{j+1}}\ne \mathbf 0$, differently from the unconstrained case (see Eq.~\ref{eq:localind}). This means that the normal matrix loses its arrow shape and local parameters of different arcs are not independent anymore.

The constrained multi-arc strategy adds great amount of information to the normal matrix $C$, as it allows the passage of information between arcs. However, several difficulties arise in the actual computations, because of the large time spans that can separate initial conditions and further dynamical effects that must be taken into account. For these reasons, pure multi-arc strategy is generally preferred to its constrained version. Nevertheless, the latter is currently used in the simulations we carry out for the BepiColombo radio-science experiment (see Sect.~\ref{sec:bc}).

\section{Interplanetary missions}
\label{sec:intermis}
In this section, we give the details about the modeling of observation and dynamical Eqs.~\eqref{eq:predict} and~\eqref{eq:dyneq} in the case of radio-science experiments for interplanetary missions. The standard scenario is a probe orbiting around or fly-bying a celestial body, while a tracking station on Earth transmits and receives radio signals to and from the spacecraft.

\subsection{Dynamics}
\label{subsec:intermis_dyn}

In order to setup the dynamical model, we must choose a suitable reference system. A common choice is to employ the so-called J2000 reference frame (also known as EME2000), which corresponds to the Equatorial reference frame defined by Earth's equator and equinox as computed at J2000 epoch. For construction, it is almost coincident with the International Celestial Reference Frame (ICRF); therefore, we will identify it as $\Sigma_{\text{ICRF}}$. Although, in some cases, it is convenient to take the solar system barycenter (SSB) as origin of the reference system, in proximity of celestial bodies it is more suitable to use a local origin. We choose the center of mass of the target body, which provides a straightforward formulation of several perturbations of the system, but it requires to add inertial forces into the dynamical model.

We describe the states of the bodies in Cartesian coordinates ($\mathbf r$,$\mathbf v$). In order to implement the system in Eq.~\eqref{eq:dyneq}, we need just to define the accelerations acting on the spacecraft, i.e. the last three components of $\mathbf f$. In the next paragraphs, we will present an exhaustive list of effects acting on a spacecraft and we will evaluate their magnitude in the case of the Juno mission. When a force is conservative, we prefer to write down its potential $U$ (per unit of probe's mass), so that the corresponding acceleration is just $\mathbf a=\nabla U$.

\begin{figure}[t]
\centering
\includegraphics[scale=0.65]{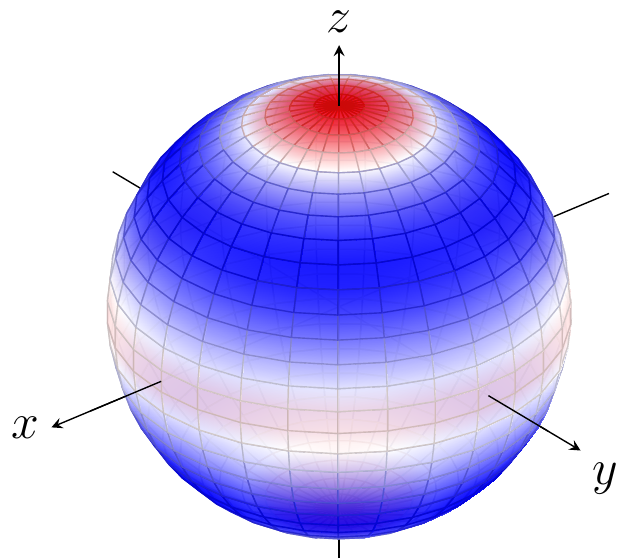}\quad
\includegraphics[scale=0.65]{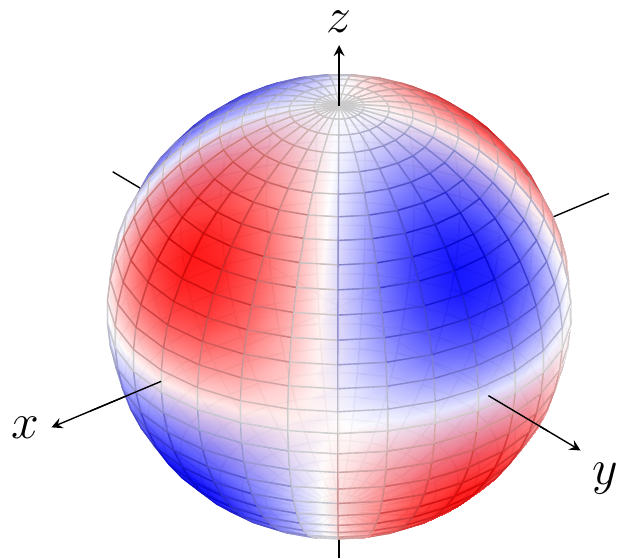}
\caption{Examples of spherical harmonics: on the left, the zonal harmonic of degree $\ell=4$ and order $m=0$; on the right, the tesseral harmonic of degree $\ell=3$ and order $m=2$. On the spheres, we reported positive values in red (positive gravitational anomaly), negative values in blue (negative gravitational anomaly) and null values in white.}
\label{fig:spherharm}
\end{figure}

We consider a spacecraft in the proximity of a target (or primary) body. The main force acting on the probe is the gravitational attraction of the body, modeled as a point mass, whose gravitational parameter is $Gm_0$ (gravitational constant per mass). The associated potential is 
\begin{equation}
\label{eq:kep}
U_0(\mathbf r)=\frac{Gm_0}{r},
\end{equation}
where $r=\abs{\mathbf r}$ and $\mathbf r$ is the position of the spacecraft with respect the center of mass of the target body (which is the origin of our reference system). $U_0$ is the potential associated to the well-known Kepler's problem: under the effect of $U_0$, the spacecraft moves along a conic (an ellipse in case of negative energy).

However, in the context we are analyzing, the number of perturbations is large. Since other bodies, like planets or satellites, are present in the system, we have to consider their gravitational effect on the probe. It is called third-body perturbation and its potential is
\begin{equation}
\label{eq:3body}
U_{3b}(\mathbf r)=\sum_j Gm_j\left(\frac{1}{\abs{\mathbf r-\mathbf r_j}}-\frac{\mathbf r\cdot\mathbf r_j}{r^3_j}\right),
\end{equation}
where the sum is over all the bodies (different from the primary) we include in the model, $Gm_j$ is their gravitational parameter and $\mathbf r_j$ their position. The third-body perturbation consists in the difference of two terms: the first is due to the direct gravitational force on the spacecraft, while the second is indirect and it is due to the attraction that the body $j$ acts on the target body.

\begin{figure}[t]
\centering
\includegraphics[scale=0.6]{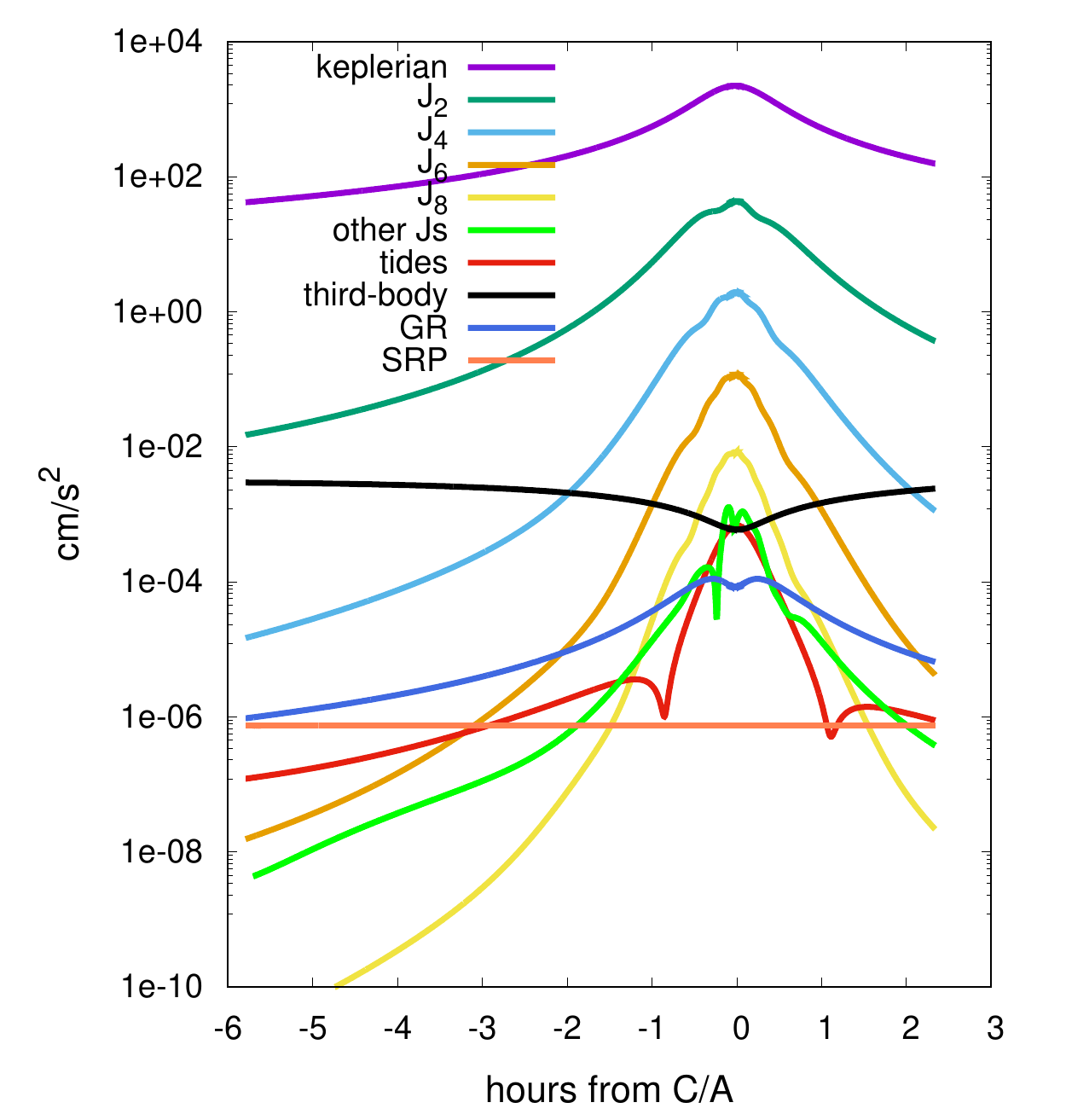}
\caption{Accelerations acting on the Juno spacecraft during PJ03. Because of the great accuracy of the data, we need to take into account effects up to about $10^{-6}$ cm/s$^2$ in order to fit the data.}
\label{fig:pj3acc}
\end{figure}

Both in Eqs.~\eqref{eq:kep} and~\eqref{eq:3body}, we have considered bodies approximated as point masses. However, since the spacecraft is close to the target body, the deviation of its mass distribution from a uniform sphere is the source of significant perturbations. The corresponding potential can be expanded in spherical harmonics series \citep{KAULA_1966} that, in a body-fixed reference frame $\Sigma_{\text{BF}}$, reads
\begin{equation}
\label{gravexp}
U_{sh}(\mathbf r^{\text{BF}})=\frac{Gm_0}{r}\sum_{\ell=2}^{\ell_{max}}\left(\frac{R_0}{r}\right)^\ell\sum_{m=0}^\ell P_{\ell m}(\sin(\phi))\left[C_{\ell m}\cos(m\theta)+S_{\ell m}\sin(m\theta)\right],
\end{equation}
where $R_0$ is the radius of the target body, $P_{\ell m}$ are the so-called Legendre associated functions of degree $\ell$ and order $m$, and $(r,\theta,\phi)$ is the representation in spherical coordinates (distance, longitude, latitude) of the position of the spacecraft $\mathbf r^{\text{BF}}$ in this new frame. Along with the gravitational parameter $Gm_0$, the spherical harmonic coefficients $C_{\ell m}$ and $S_{\ell m}$ describe the whole static gravity field of the primary body. The parameters $J_\ell=-C_{\ell 0}$ are called zonal coefficients, while the parameters with $m\ne 0$ are called tesseral (see \figurename~\ref{fig:spherharm}). In particular, $J_2$ is the main gravitational anomaly, as it is associated with the deformation caused by the spin of the body around its axis, and it is the source of one of the largest perturbations in space dynamics (see \figurename~\ref{fig:pj3acc}).  

In order to compute the acceleration associated with Eq.~\eqref{gravexp} in the reference frame $\Sigma_{\text{ICRF}}$ that we use to propagate the dynamics, we must perform a suitable rotation. We define the matrix $\mathcal R$ at the time $t$ as
\begin{equation}
\label{eq:bfrot}
\mathcal R(t) =  \mathcal R_3(\psi(t))\mathcal R_1(\pi/2-\delta(t))\mathcal R_3(\alpha(t)+\pi/2),
\end{equation}
where $\mathcal R_i(\mathcal X)$ is the rotation matrix of an angle $\mathcal X$ around the $i$th axis. The angles $\alpha$ and $\delta$ are the right ascension and the declination of body's north pole in $\Sigma_{\text{ICRF}}$, while $\psi$ is the angle that defines the position of body's prime meridian along its equator. They depend on time, because of body's spin and the precession of its axis. For planets and satellites, they can be retrieved from models developed in literature, such as the ones reported in \citet{ARCHINAL-etal_2018}, or must be propagated with a suitable dynamics (see, e.g., \citealt{FOLKNER-etal_2014}). Finally, as $\mathbf a_{sh}^{\text{BF}}=\nabla U_{sh}(\mathbf r^{\text{BF}})$ is the acceleration in the body-fixed reference frame, the acceleration in $\Sigma_{\text{ICRF}}$ is $\mathbf a_{sh}=\mathcal R\mathbf a_{sh}^{\text{BF}}$.

Another important perturbation of gravitational nature is the one associated with the tides raised on the primary by other bodies. While the expression in Eq.~\eqref{gravexp} describes the static gravity anomalies of the target body, tidal deformation changes with time, because of the motion of the bodies that generate it. Tidal potential is commonly expanded in series by means of the Love numbers $k_\ell$ \citep{LOVE_1909}:
\begin{equation}
U_{tid}(\mathbf r)=\sum_j\frac{Gm_j}{r_j}\sum_{\ell=2}^{+\infty}k_\ell\left(\frac{R_0}{r}\right)^{\ell+1}\left(\frac{R_0}{r_j}\right)^\ell P_\ell(\cos\gamma_j),
\end{equation}
where $P_{\ell}$ are the so-called Legendre polynomials of degree $\ell$ and $\gamma_j$ is the angle between the position of the spacecraft $\mathbf r$ and the external body $\mathbf r_j$. Coefficients $k_\ell$ describe the response of the gravitational field of the primary due to the deformation induced by the gravitational attraction of external bodies. Otherwise, the tidal contribution can directly added to the spherical harmonic coefficients $C_{\ell m}$ and $S_{\ell m}$:
\begin{equation}
\begin{aligned}
\Delta C_{\ell m} = & \frac{k_{\ell m}}{2\ell +1}\sum_j\frac{m_j}{m_0}\left(\frac{R_0}{r_j}\right)^{\ell+1} P_{\ell m}(\sin\phi_j) \cos(m\lambda_j),\\
\Delta S_{\ell m} = & \frac{k_{\ell m}}{2\ell +1}\sum_j\frac{m_j}{m_0}\left(\frac{R_0}{r_j}\right)^{\ell+1} P_{\ell m}(\sin\phi_j) \sin(m\lambda_j),
\end{aligned}
\end{equation}
where $k_{\ell m}$ are the new tidal parameters and $(r_j,\theta_j,\phi_j)$ are the spherical coordinates of the bodies that raise the tides in $\Sigma_{\text{BF}}$.

\begin{table}[t]
{\small
\begin{center}
\begin{tabular}{|l|l|c|c|}
\hline
Parameter  &  Effect on the metric  & GR & BC\\
\hline
$\gamma$   & measure of the curvature of space-time & 1 & *\\
\hline
$\beta$    & measure of the non-linearity of the superimposition law for gravity & 1 & *\\
\hline
$\xi$      & measure of preferred-location effects & 0 & *\\
\hline
$\alpha_1$, $\alpha_2$ & measure of preferred-frame effects & 0 & *\\
\hline
$\alpha_3$ & measure of preferred-frame effects \& & 0 & \\
           & measure of the violation of the law of conservation of momentum &  & \\

\hline
$\zeta_1$, $\zeta_2$, $\zeta_3$, $\zeta_4$ & measure of the violation of the law of conservation of momentum & 0 & \\
\hline
\end{tabular}
\end{center}
}
\caption{List of the post-Newtonian parameters as presented in \citet{WILL_2014} along with their effect on the metric and their value in the General Relativity theory. The asterisks indicate the parameters that will be estimated with the BepiColombo relativity experiment \citep{SCHETTINO-etal_2018}; the others are necessarily equal to $0$ in every semi-conservative theory of gravity. Actually, in place of the parameter $\xi$, we can equivalently solve for the Nordtvedt parameter $\eta$ (see Sect.~\ref{sec:bc}), which measures possible violations of the strong equivalence principle \citep{WILL_2014}.}
\label{tab:pntab}
\end{table}

The proximity to a massive target body, like a planet, can make the Newtonian formulation of the gravity not accurate enough. In order to take into account the deviation described in the General Relativity framework, the dynamical model must include the acceleration
\begin{equation}
\label{eq:genrel}
\mathbf a_{gr}(\mathbf r,\mathbf v)=\frac{Gm_0}{c^2r^3}\left[\left(2(\gamma+\beta)\frac{Gm_0}{r}-v^2\right)\mathbf r+2(1+\gamma)(\mathbf r\cdot \mathbf v)\mathbf v\right],
\end{equation}
where $\mathbf v$ is the velocity of the probe, $v=\abs{\mathbf v}$ and $c$ the light speed. In Eq.~\eqref{eq:genrel}, $\gamma$ and $\beta$ are two post-Newtonian parameters (see, e.g., \citealt{WILL_2014}), that in General Relativity are equal to $1$. Other exotic relativistic effects can intervene in the dynamics, although they can be significant only for the motion of bodies close to the Sun (like Mercury). In \tablename~\ref{tab:pntab}, we reported all the post-Newtonian parameters with their contribution to the metric of the space-time. For a detailed description of their dynamical effects, we refer to \citet{MILANI-etal_2010}.

Apart from the effects described above, there are other minor gravitational and relativistic forces that can be worth including into the dynamical model (e.g., geodetic precession, Lense-Thirring effect and so on). Moreover, other inertial forces due to the choice of the origin appear because of the gravitational attraction of the external bodies on the non-spherical field of the primary body, such as the so-called indirect oblateness relative to the $J_2$ coefficient. In order to investigate if some of these effects have a significant impact in the dynamics of the spacecraft and in the OD experiment, a sensitivity analysis must be carried out (see, e.g., \citealt{TOMMEI-etal_2015}).

For bodies with an high area-to-mass ratio $A/m$ like a spacecraft, also non-gravitational forces must be included into the dynamics \citep{MILANI-etal_1987}. These forces arise from sources of different nature than the presence of a mass in space, like the radiation from the Sun. Indeed, the photons hitting the spacecraft at the speed of light transfer a small momentum to the probe, which generates an acceleration called solar radiation pressure (SRP):
\begin{equation}
\mathbf a_{srp}=-C_{srp}\frac{\Phi_{\astrosun}}{c}\frac{A}{m}\cos{(b)}\mathbf{\hat s},
\end{equation}
where $b$ is the angle between the normal $\mathbf{\hat n}$ to the hit surface and the direction of the Sun $\mathbf{\hat s}$, and $\Phi_{\astrosun}$ is the solar radiation flux at the position of the spacecraft. $C_{srp}$ is a parameter whose value depends on the material of the hit surface and is generally close to $1$. For very precise modeling of the solar radiation pressure, it is necessary a realistic shape model of the spacecraft, taking into account all its components and their physical properties (see, e.g., \citealt{ZIEBART_2004}).

Other non-gravitational forces are the indirect radiation pressure due to the radiation reflected by the target body, space maneuvers and so on. In particular, space maneuvers are generally used far from the periods in which observations are collected; therefore, even if they produce a significant acceleration, they can be neglected in the dynamical model used for the OD experiment (see Sect.~\ref{subsec:adv_mult}).

In \figurename~\ref{fig:pj3acc}, we reported the magnitude of the forces acting on the Juno spacecraft during PJ03. As it can be noted, the main perturbations are due to the even zonal harmonics of the gas giant.

As showed in Eq.~\eqref{eq:parder}, along with the equations of motion in Eq.~\eqref{eq:dyneq}, we must integrate also the differential equations of the partial derivatives with respect to the dynamical parameters and the state of the spacecraft. In Orbit14, we compute analytically the partial derivatives of the accelerations, $\partial \mathbf f/\partial p$ and $\partial \mathbf f/\partial \mathbf y$, in order to avoid possible issues in the numerical accuracy of the finite differences computation.

While the propagation of the spacecraft is necessary to perform radio-science experiments, when the experiment is not sensitive to the position and velocity of other bodies, we can take them from ephemerides. There are tables from which we can retrieve the states of celestial bodies at the times needed in the computations. Ephemerides are provided, for instance, by JPL\footnote{https://naif.jpl.nasa.gov/pub/naif/generic\_kernels} and the Institut de Mécanique Céleste et de Calcul des \'Ephémérides (IMCCE)\footnote{https://www.imcce.fr/services/ephemerides}, and can be handled in the software through the SPICE Toolkit. Also the orientation of the Earth, which is essential to know the position of the tracking antenna in $\Sigma_{\text{ICRF}}$, can be obtained from Earth Orientation Parameters provided by the International Earth Rotation and Reference Systems Service (IERS)\footnote{https://hpiers.obspm.fr/eop-pc}. 

However, in some cases, we cannot just propagate the spacecraft, but we need to integrate also other bodies. This happens when we want to fit parameters that affect the dynamics of the target body or we want to improve the knowledge of its orbit. For example, for the BepiColombo radio-science experiment, Mercury is propagated to investigate possible deviations from the General Relativity through the fit of its orbit \citep{MILANI-etal_2010}. Moreover, multiple fly-bys mission like Cassini and JUICE are excellent sources of data to compute accurate orbits of planet's satellites and study dynamical effects such as tidal dissipation \citep{DIRKX-etal_2017,LAINEY-etal_2020}.

Therefore, sometimes, we must include other dynamics in Eq.~\eqref{eq:dyneq} than that of the spacecraft and add new parameters in $\mathbf y_0$. Typical dynamics we are interested in are: motion of the planets, motion of the satellites of a planet and rotational dynamics of a target body. They can be affected by new parameters that we did not consider in the motion of the spacecraft. Although the new dynamical parameters do not intervene directly in the dynamics of the probe, they have an indirect effect through the state of other celestial bodies. Because of the limited time span covered by space missions and the scarce influence that small changes of certain dynamics have on the others, in first approximation, we can integrate the dynamics according to the following order:
\begin{itemize}
\item motion of the barycenters of the planets (using satellites' ephemerides);
\item motion of the satellites (using existing models of the planets' orientation);
\item orientation of the planets and satellites;
\item motion of the spacecraft (which does not influence at all other dynamics).
\end{itemize}

For example, let us assume that we want to estimate a dynamical parameter $p$ that influences the motion of a celestial body B involved in spacecraft's dynamics. First, we must propagate the state $\mathbf y^{\text{B}}=(\mathbf r^{\text{B}},\mathbf v^{\text{B}})$ and its partial derivative $\partial\mathbf y^{\text{B}}/\partial p$. Then, when we integrate the dynamics of the spacecraft, for computing its acceleration at time $t$, we must take the state of the body $\mathbf y^{\text{B}}(t)$ from our previous integration. Following Eq.~\eqref{eq:parder}, the differential equation for $\partial\mathbf y/\partial p$ is
\begin{equation}
\label{eq:parderchain}
\frac{d}{dt}\frac{\partial\mathbf y}{\partial p}=\frac{\partial \mathbf f}{\partial \mathbf {y}}\frac{\partial \mathbf {y}}{\partial \mathbf y^{\text{B}}}\frac{\partial\mathbf y^{\text{B}}}{\partial p},
\end{equation}
where, again, $\partial\mathbf y^{\text{B}}(t)/\partial p$ is retrieved from the previous integration.

%Such scheme is useful, for example, also for estimating the polar moment of inertia (MoI) of the target body. In this case, $\mathbf y^{\text{B}}$ is the vector $(\boldsymbol \alpha^{\text{B}}, \dot{\boldsymbol\alpha}^{\text{B}})$ of the angles that define the orientation of the planet (e.g., its Euler angles or the ones presented in Eq.~\ref{eq:bfrot}) and their first derivatives, while $p$ is the MoI, which affects the precession of the body's spin axis. The vector of state is propagated through the equations of motion of a rigid body with the inclusion of the external torques of the Sun, planets and satellites acting on the oblate shape of the target body. The state $\mathbf y^{\text{B}}(t)$ intervenes in the spacecraft's dynamics mainly through the acceleration in Eq.~\eqref{gravexp}. Therefore, the computation of the differential equation for $\partial\mathbf y/\partial\text{MoI}$ must follow the chain rule described in Eq.~\eqref{eq:parderchain}.

\begin{figure}[t]
\centering
\includegraphics[scale=0.6]{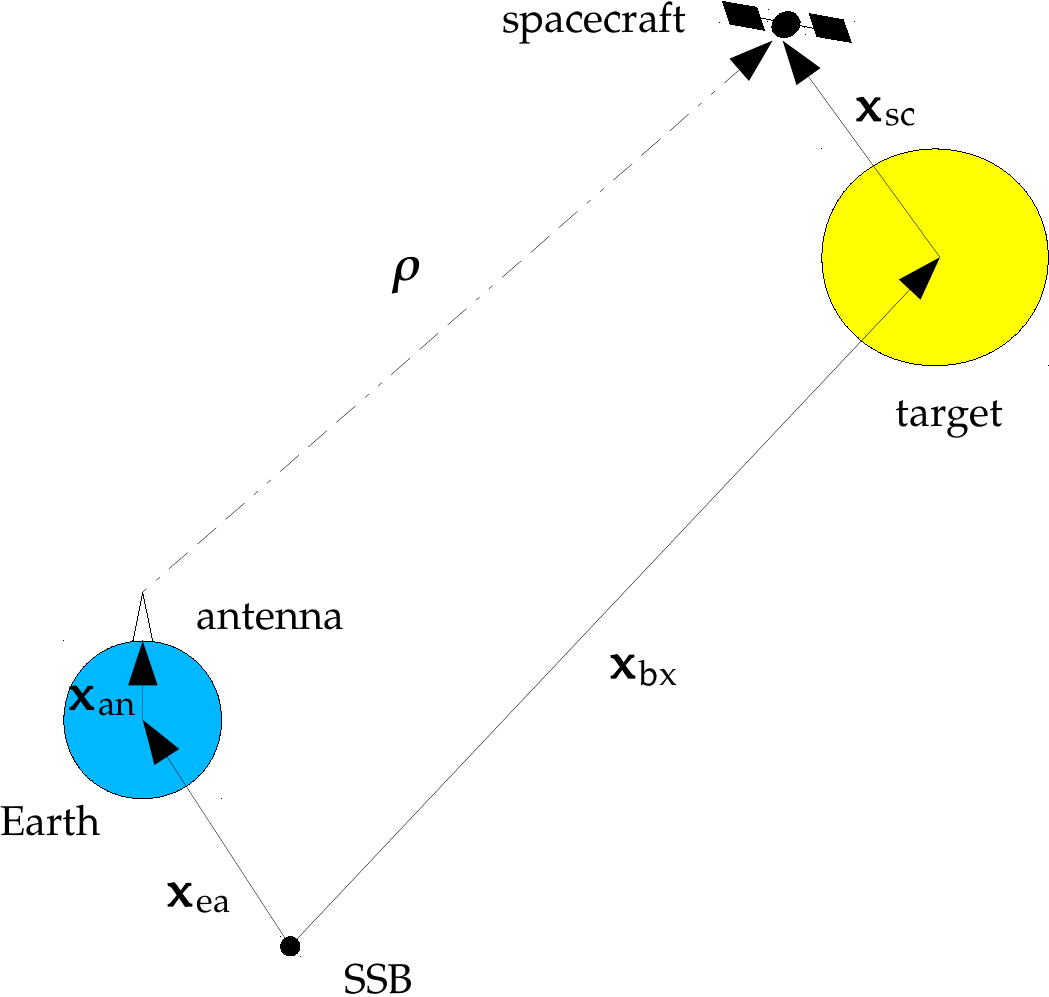}
\caption{General configuration of an interplanetary mission and distance vector $\boldsymbol\rho$ between the tracking antenna and the spacecraft.}
\label{fig:radioscheme1}
\end{figure}

\subsection{Computed observables}
\label{subsec:intermis_obs}
In Sect.~\ref{subsec:intermis_dyn}, we have described the dynamical model (Eq.~\ref{eq:dyneq}) of a spacecraft, now we need to define the observation function in Eq.~\eqref{eq:predict}. For radio-science experiments we have two kinds of data: range and Doppler, or, equivalently, range and range-rate. Range data give a measure of the 2-way light-time travel $\tau$ of the radio signal from the tracking station to the spacecraft and return. As the signal moves at the speed of light, $\tau$ is related to the distance $\rho$ of the spacecraft from the station through $\tau=2\rho/c$. Doppler data measures the change in frequency of the signal from transmission to reception. Indeed, the variation of the distance of the spacecraft $\dot\rho$ (range-rate), due to its large speed and to the dynamical effects that it undergoes, is reflected in a Doppler effect in the final received signal.

In order to predict the values of these observables, we can follow the scheme presented in \figurename~\ref{fig:radioscheme1}. The distance vector $\boldsymbol \rho$ can be expressed as the difference between the position of the spacecraft and the antenna in a reference frame centered in the solar system barycenter. The vector of the spacecraft in such a frame is decomposed as the sum of the vector from SSB to the center of mass of the target body $\mathbf x_{bx}$ and the vector from the center of the target body to the spacecraft $\mathbf x_{sc}$, following the structure of the dynamics described in Sect.~\ref{subsec:intermis_dyn}. In the same way, the vector of the antenna is decomposed as the sum of the vector from SSB to the center of mass of the Earth $\mathbf x_{ea}$ and the vector from the center of the Earth to the antenna $\mathbf x_{an}$. The scheme in \figurename~\ref{fig:radioscheme1} is the most simple one for an interplanetary mission orbiting around a celestial body, but more complicated configurations can occur (see, e.g., \citealt{MILANI-etal_2002} and \citealt{LARI-MILANI_2019}).

\begin{figure}[t]
\centering
\includegraphics[scale=0.6]{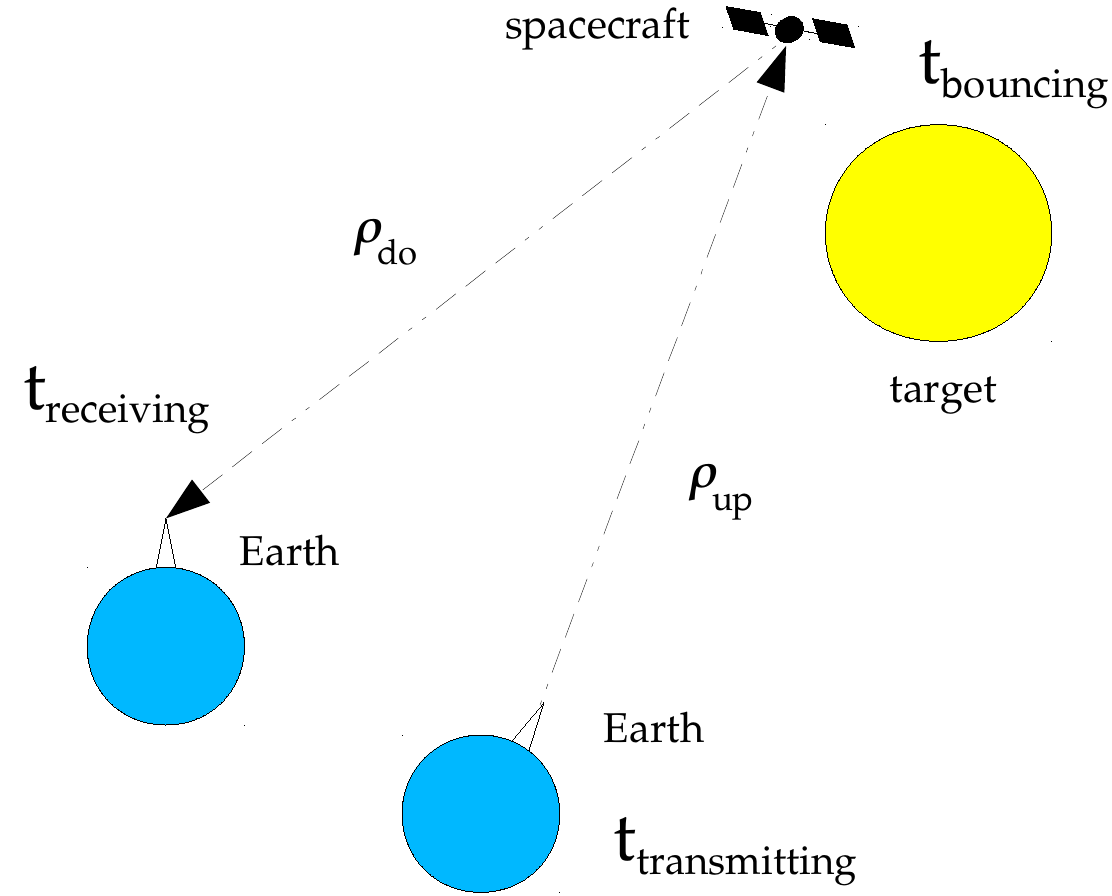}
\caption{Schematic representation of the three different times considered in the computation of the range (time of transmission, bouncing and reception).}
\label{fig:radioscheme2}
\end{figure}

In the end, the distance we want to model is given by
\begin{equation}
\label{range}
\tilde\rho=\abs{\boldsymbol\rho}+S(\gamma)=\abs{(\mathbf x_{bx}+\mathbf x_{sc})-(\mathbf x_{ea}+\mathbf x_{an})}+S(\gamma),
\end{equation}
where the term $S(\gamma)$ is the Shapiro effect \citep{SHAPIRO_1964}
\begin{equation}
\label{shapiro}
S(\gamma)=\frac{(1+\gamma)Gm_{\astrosun}}{c^2}\ln{\left(\frac{r_t+r_r+r}{r_t+r_r-r}\right)},
\end{equation}
which accounts for the curvature of the space-time due to the Sun. If the radio signal passes in the proximity of other massive bodies (such as the target one), it could be necessary to add also their effects in Eq.~\eqref{shapiro}. The Shapiro effect depends on the post-Newtonian parameter $\gamma$, while $r_r$ and $r_t$ are the distances of the receiver and the transmitter from the Sun, and $r$ their mutual distance.

However, Eq.~\eqref{range} is not complete, as it lacks the times at which the vectors are computed. Indeed, radio signal is transmitted from the tracking station on Earth at time $t_t$, it reaches the spacecraft that sends it back at time $t_b$ (bouncing time) after changing the frequency of a factor $\alpha$ (turnaround ratio) and it is received by the ground station at time $t_r$. As the signal covers millions of kilometers, the three times can differ for tens of minutes or even hours. Therefore, when we compute the trip of the radio signal to predict the data value, we have to consider that the Earth moved in this time span, as showed in \figurename~\ref{fig:radioscheme2}. Moreover, when the signal arrives and the data point is collected, we know the receiving time $t_r$, but not the corresponding $t_b$ and $t_t$.

Therefore, we need an algorithm to recover these remaining two times starting from $t_r$. We consider the downlink trip, and we rewrite Eq.~\eqref{range} as:
\begin{equation}
\label{rangedown}
\tilde\rho_{do}=\abs{\boldsymbol{\rho}_{do}}+S_{do}=\abs{(\mathbf x_{bx}(t_b)+\mathbf x_{sc}(t_b))-(\mathbf x_{ea}(t_r)+\mathbf x_{an}(t_r))}+S_{do}(\gamma),
\end{equation}
where the first two vectors on the right hand side are computed at the bouncing time, while the last two are computed at the receiving time. The quantity $S_{do}(\gamma)$ is the Shapiro effect on the down path of the signal.

In order to recover $t_b$, we perform an iterative process. We start from a first guess given by taking the positions of the Earth and of the target body at time $t_r$. Then, we compute a first estimation of the downlink range presented in Eq.~\eqref{rangedown}, from which we can compute a new value for the travel time $\tau_{do}=\tilde\rho_{do}/c$ of the signal. We update the value of $t_b$ with $t_r-\tau_{do}$. At the following iteration, we use the new value and we remake the computation, continuing until the variation in $t_b$ is below a certain threshold (e.g., $10^{-12}$ seconds). More details of this procedure can be found in \citet{TOMMEI-etal_2010}.

Once $t_b$ is computed, we consider the uplink trip
\begin{equation}
\label{rangeup}
\tilde\rho_{up}=\abs{\boldsymbol{\rho}_{up}}+S_{up}=\abs{(\mathbf x_{bx}(t_b)+\mathbf x_{sc}(t_b))-(\mathbf x_{ea}(t_t)+\mathbf x_{an}(t_t))}+S_{up}(\gamma),
\end{equation}
where, this time, the positions of the Earth and of the antenna are taken at the transmission time. In order to compute $t_t$, we have to perform a similar iterative process to the one used for $t_b$. We can start with $t_t=t_b+\tau_{do}$ as a first guess (light-time travel in uplink is similar, but not equal to the one in downlink), and using Eq.~\eqref{rangeup}, we iterate the computation until we obtain the final value of $t_t$.

\begin{figure}[t]
\includegraphics[scale=0.45]{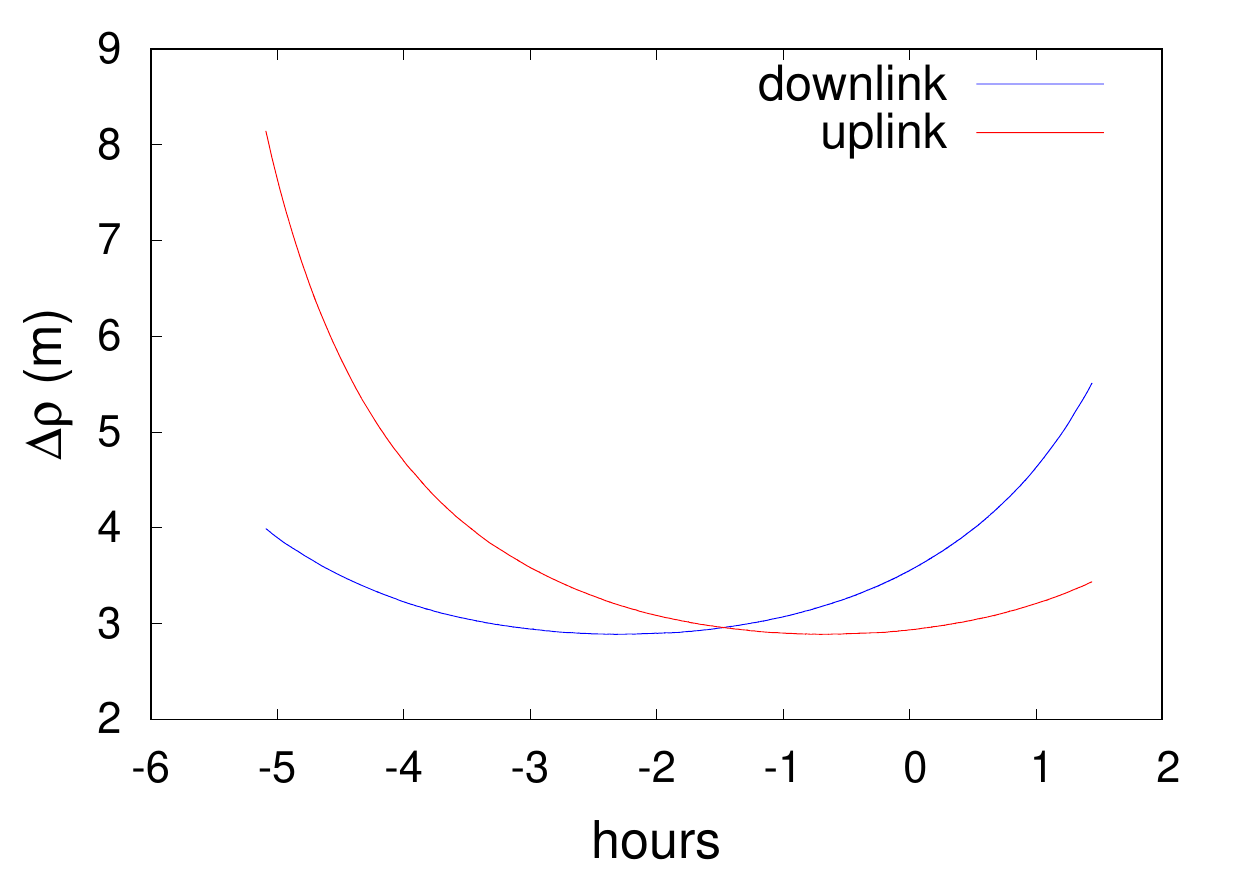}
\includegraphics[scale=0.45]{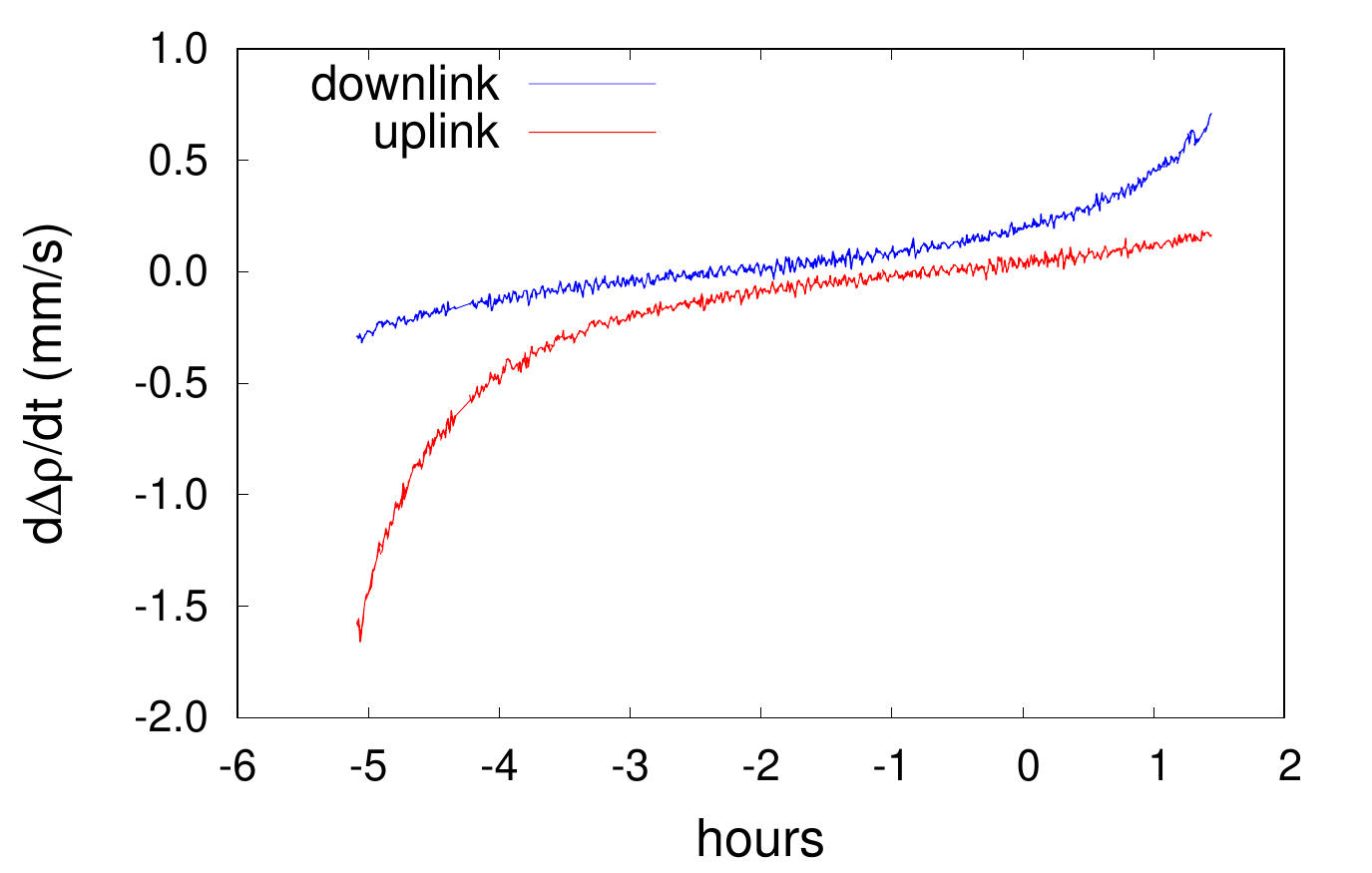}
\caption{Contribution of the troposphere on range data (left) and range-rate (right) for Juno's PJ03. These effects have a signal-to-noise ratio between $10$ and $100$, therefore they cannot be neglected in the processing of the data.}
\label{fig:pj3tropo}
\end{figure}

In the end, the total 2-way light-time travel is given by $\tau=t_r-t_t$ and the final formulation of the 2-way range is
\begin{equation}
\label{finalrange}
\tilde\rho_{2w}=\tilde\rho_{do}+\Delta\tilde\rho_{do}+\tilde\rho_{up}+\Delta\tilde\rho_{up},
\end{equation}
where $\Delta\tilde\rho_{do}$ and $\Delta\tilde\rho_{up}$ accounts for the delay due to troposphere and ionosphere, and eventually other exotic effects experienced by the signal along its path (see, e.g., \citealt{ZANNONI-etal_2019} about the Io plasma torus). In particular, the interaction between the radio signal and the Earth's troposphere produces a significant systematic signal in the observations (see \figurename~\ref{fig:pj3tropo}), mostly due to the presence of water vapor molecules \citep{ASMAR-etal_2005}. For a correct modeling of this effect, it is necessary to have measurements of water vapor over the tracking station, both in transmission and reception. Such measurements can be generically obtained from the atmospheric GPS data, or from high-precision instruments, like the Advanced Water Vapor Radiometer used also during the collection of Juno radio-science data \citep{ASMAR-etal_2017}.

In order to compute the range-rate, we derive Eq.~\eqref{finalrange} with respect to time. As the expression for the range contains three different times, we choose to derive with respect to $t_r$. For the downlink we have
\begin{equation}
\label{derivrangedown}
\frac{d \tilde\rho_{do}}{d t_r}=\frac{\boldsymbol{\rho}_{do}}{\abs{\boldsymbol{\rho}_{do}}}\cdot\left(\frac{\partial(\mathbf x_{bx}(t_b)+\mathbf x_{sc}(t_b))}{\partial t_b}\frac{d t_b}{d t_r}-\frac{\partial(\mathbf x_{ea}(t_r)+\mathbf x_{an}(t_r))}{\partial t_r}\right)+\frac{dS_{do}}{d t_r}
\end{equation}
where vectors $\partial \mathbf y(t)/\partial t$ are the velocities of the bodies at time $t$ and
\begin{equation}
\label{derivtimetb}
\frac{d t_b}{d t_r}=1-\frac{1}{c}\frac{d \tilde\rho_{do}}{d t_r}.
\end{equation}
In the same way, for the uplink
\begin{equation}
\label{derivrangeup}
\frac{d \tilde\rho_{up}}{d t_r}=\frac{\boldsymbol{\rho}_{up}}{\abs{\boldsymbol{\rho}_{up}}}\cdot\left(\frac{\partial(\mathbf x_{bx}(t_b)+\mathbf x_{sc}(t_b))}{\partial t_b}\frac{d t_b}{d t_r}-\frac{\partial(\mathbf x_{ea}(t_t)+\mathbf x_{an}(t_t))}{\partial t_t}\frac{d t_t}{d t_r}\right)+\frac{dS_{up}}{d t_r}
\end{equation}
where
\begin{equation}
\label{derivtimett}
\frac{d t_t}{d t_r}=1-\frac{1}{c}\frac{d \tilde\rho_{up}}{d t_r}.
\end{equation}

Since the derivatives of the times in Eqs.~\eqref{derivtimetb} and~\eqref{derivtimett} contain the quantities we want to compute, the expressions in Eqs.~\eqref{derivrangedown} and~\eqref{derivrangeup} are not explicit. Therefore, also for the range-rate, we use an iterative procedure. We start from a first guess of $d \tilde\rho_{do}/d t_r$ approximating $d t_b/d t_r=1$ and we iterate up to convergence. Similarly, we compute $d \tilde\rho_{up}/d t_r$, so that the final formulation of the range-rate (2-way) is
\begin{equation}
\label{finalrangerate}
\frac{d \tilde\rho_{2w}}{d t_r}=\frac{d \tilde\rho_{do}}{d t_r}+\frac{d \Delta\tilde\rho_{do}}{d t_r}+\frac{d \tilde\rho_{up}}{d t_r}+\frac{d\Delta\tilde\rho_{up}}{d t_r}.
\end{equation}
For the complete formulas of $dS_{do}/d t_r$ in Eq.~\eqref{derivrangedown} and $dS_{up}/d t_r$ in Eq.~\eqref{derivrangeup}, we refer to \citet{TOMMEI-etal_2010}.

Eq.~\eqref{finalrangerate} describes the instantaneous value of the range-rate. However, the actual observation is not an instantaneous measure of the frequency of the radio signal, but it is obtained through an integration over time. Therefore, also the computed data must account for this averaging of the measure. If we assume an integration time $\Delta$, which generally goes from $10$ to $1000$ s, we have
\begin{equation}
\label{finalrangerateinteg}
\overline{\frac{d \tilde\rho_{2w}}{d t_r}}=\int_{t_r-\Delta/2}^{t_r+\Delta/2}\frac{d \tilde\rho_{2w}(s)}{d s}ds
\end{equation}

As range and range-rate are equivalent to the real observations given in terms of the properties of the received signal, for simulations of the radio-science experiments we can use them. Indeed, covariance analysis is not affected in any way by this choice.

However, in the case we must process real data, we need the corresponding formulation of Eqs.~\eqref{finalrange} and~\eqref{finalrangerateinteg} in terms of received phase and frequency of the signal. In particular, Doppler data are defined as the received frequency at the tracking station; the total Doppler effect can be computed as the difference between the received and transmitted frequency, once we have accounted for the modulation of the signal at the spacecraft.

\begin{figure}[t]
\centering
\includegraphics[scale=0.45]{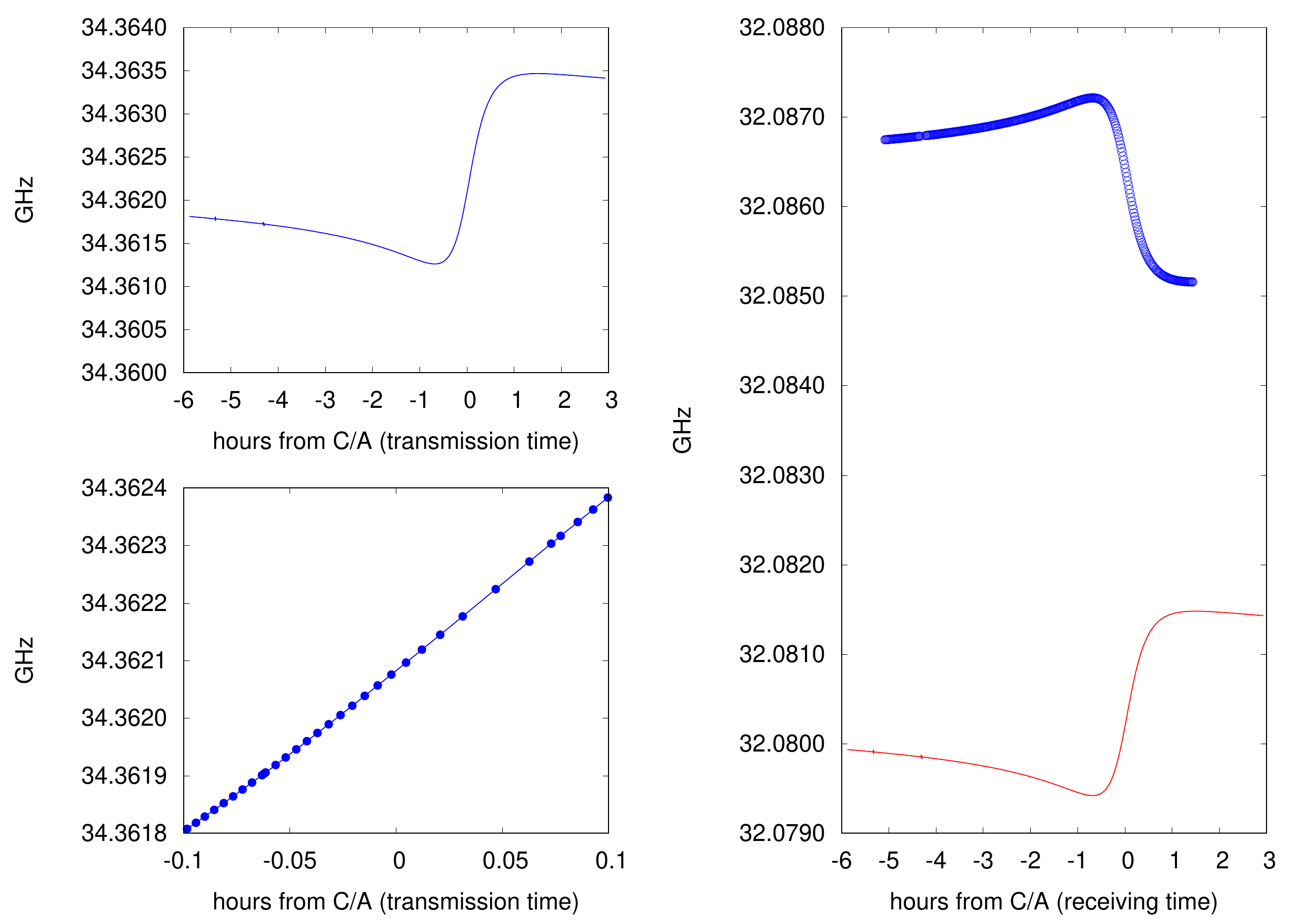}
\caption{Transmitted (left) and received (right) frequencies at Juno's PJ03. On the bottom left, we zoomed the plot on the top, in order to show the ramps' intervals near the pericenter of Juno's orbit. On the right, we reported the real received frequencies in blue, while in red we draw the signal without the Doppler effect and accounting only for the frequency modulation at the spacecraft ($\alpha=3360/3599$).}
\label{fig:signal}
\end{figure}

The transmitted frequency is not necessarily constant, but it can be ramped, as shown in \figurename~\ref{fig:signal}. Ramps make the frequency change and are designed in such a way that the total frequency slope in the data is approximately equal to the frequency slope in transmission. The whole period of transmission from the tracking station can be subdivided in intervals $[T_0^i,T_0^{i+1}]$ for which the frequency of the transmitted signal is
\begin{equation}
\label{ramp}
f(t)=f_0^i+\dot f^i (t-T_0^i), \quad t\in[T_0^i,T_0^{i+1}],
\end{equation}
where $f_0^i$ is the base value of the $i$th ramp and $\dot f^i$ is the constant rate of the ramp (see \figurename~\ref{fig:signal}).

\citet{MOYER_2003} gives an exhaustive description of the formulas for computing radio-science data. In the case of a 2-way ramped signal, Doppler data at the receiving time $t_r$ (integration time $\Delta$) are obtained integrating the transmitted frequency:
\begin{equation}
\label{eq:dopplermoyer}
\mathcal D(t_r)=\frac{\alpha}{\Delta}\int_{\Delta'(t_t)}f(t)dt, \quad t_t=t_r-\tilde\rho_{2w}(t_r)/c;
\end{equation}
where $\alpha$ is the modulation factor operated by the spacecraft (see \figurename~\ref{fig:signal}) and $\Delta'(t_t)$ is the integration interval at the transmission time corresponding to $\Delta$, through the light-time travel $\tilde\rho_{2w}/c$. In our implementation, we use a slightly different formulation, considering the integral in Eq.~\eqref{eq:dopplermoyer} in the receiving time dominion
\begin{equation}
\label{eq:dopplermoyerinv}
\mathcal D(t_r)=\frac{\alpha}{\Delta}\int_{t_r-\frac{\Delta}{2}}^{t_r+\frac{\Delta}{2}}f(t')\left[1-\frac{1}{c}\frac{d\tilde\rho_{2w}(t)}{dt}\right]dt, \quad t'=t-\tilde\rho_{2w}(t)/c.
\end{equation}
In this way, we have an explicit dependence on the range-rate. The integral in Eq.~\eqref{eq:dopplermoyerinv} is computed numerically; if in the integration interval $[t_r-\Delta/2,t_r+\Delta/2]$ there is one or more changes of ramp, we subdivide it in sub-intervals according to Eq.~\eqref{ramp} and we compute the corresponding integrals separately.

Range data are defined as the difference in phase between the received and transmitted signal. They are given in range units (RU), which corresponds to two cycles of the signal frequency. The formulation for range data as given by \citet{MOYER_2003} is
\begin{equation}
\label{eq:rangemoyer}
\mathcal R(t_r)=\alpha\int_{t_t}^{t_r}f(t)dt \quad \text{(mod $M$)}, \quad t_t=t_r-\tilde\rho_{2w}(t_r)/c;
\end{equation}
where $M$ is a fixed large value ($M\approx 10^9$ RU in the case of Juno). Differently from Eq.~\eqref{eq:dopplermoyerinv}, this integral can be computed analytically, after having subdivided it according to Eq.~\eqref{ramp}.

The uncertainties of the data depend on the frequency band. Higher frequency are more insensitive to some dispersive effects (e.g., interaction with ionosphere), reducing the noise of the observables. KaT technology provides a nominal accuracy for the Doppler of about $3\times 10^{-4}$ Hz over an integration time of $1000$ s (Ka band). For interplanetary missions, Doppler data are computed over an integration time of tens of seconds, according to the minimum timescale we are interested for the spacecraft dynamics. In order to assess the nominal uncertainty for different integration times, we can rescale assuming a Gaussian noise model $\sigma^\Delta_i=\sigma^{(1000)}\sqrt{1000/\Delta}$, which for $\Delta=60$ s results in $\sigma^{(60)}\approx 1.2\times 10^{-3}$ Hz (Ka band). In order to translate the uncertainty of the Doppler to range-rate, we can use
\begin{equation}
\label{eq:translateunc}
\sigma(\dot \rho)=\sigma(\mathcal D)\frac{f}{c},
\end{equation}
that for $f\approx 32$ GHz (received Ka frequency) and integration time of $60$ s is $\sigma(\dot \rho)\approx 1.2 \times 10^{-2}$ mm/s.

Finally, we remind that, in the OD process, we need to calculate also the partial derivatives of the computed observables with respect to the fit parameters, as described in Eq.~\eqref{eq:derres}. Computed data depend on the fit parameters through the state of the bodies involved in the definition of range and range-rate (see Eq.~\ref{range}). Because range and Doppler data as defined in Eqs.~\eqref{eq:rangemoyer} and~\eqref{eq:dopplermoyerinv} depend on $\tilde \rho$ and $d \tilde \rho/dt_r$, through a derivative chain we obtain $\partial \mathcal R/\partial \mathbf x$ and $\partial \mathcal D/\partial \mathbf x$.

\subsection{Multi-chart approach}
\label{subsec:intermis_mult}
In \figurename~\ref{fig:radioscheme1}, all the vectors are meant to be defined in the reference frame centered in SSB. The time associated with this frame is the Barycentric Dynamical Time (TDB), which is the time of planets' ephemerides. However, the states of the antenna and of the spacecraft are generally computed in a different reference system, which has its own dynamical time. Indeed, as we have described in Sect.~\ref{subsec:intermis_dyn}, for the dynamics of the spacecraft, we use a reference frame centered in the center of mass of the target body. Also the state of the antenna is computed in a local reference frame, with origin in the Earth's center of mass.

Following the relativistic formulation of \citet{KLIONER-etal_2010}, when we choose a reference frame centered in a specific body to describe the dynamics of an object, we should use a local chart with its own space-time coordinates. In order to sum vectors from different charts, as in Eq.~\eqref{range}, we must perform a transformation of coordinates from the local charts to the global chart with origin in SSB and time TDB. In particular, the antenna is given in the chart of the Earth with its Terrestrial Dynamical Time (TDT, or TT), while spacecraft's state is propagated in the chart of the target body with its TDX, where X stays for the name of the body. For example, \citet{TOMMEI-etal_2010} introduced TDM for BepiColombo dynamics around Mercury and \citet{TOMMEI-etal_2015} introduced TDJ for Juno dynamics around Jupiter.

Then, we need the expression of these space-time transformations. For the sake of generality, we consider a generic local chart around a body X, that can be also the Earth. In the next formulas, coordinates and times of the local chart are indicated with $TX$, while the ones of the global chart with $TB$.

Following \citet{TOMMEI-etal_2010}, the differential equation for the local time $t^{TX}$ in function of the SSB time $t^{TB}$ is
\begin{equation}
\label{eq:changetime}
\frac{dt^{TX}}{dt^{TB}}=1-\frac{1}{c^2}\left(U+\frac{v_X^2}{2}-L_X\right),
\end{equation}
where $c$ is the light speed, $U$ is the gravitational potential at the center of the body X due to the Sun and other bodies and $v_X=\abs{\mathbf v_X}$ is the norm of the velocity of the body X in the SSB chart. The constant $L_X$ is a scaling factor used to remove secular terms in the difference between the definition of the two times (see, e.g, \citealt{TURYSHEV-etal_2013}). In order to get the time transformation between $t^{TX}$ and $t^{TB}$, the differential Eq.~\eqref{eq:changetime} must be integrated for the whole time span we consider for the radio-science experiment (nominally, the duration of the space mission). In literature, there are also some approximate analytical solutions of Eq.~\eqref{eq:changetime} (see \citealt{FAIRHEAD-etal_1988}), which can be used to perform the time transformations.

The transformation of the space coordinates $(\mathbf x, \mathbf v)$ is given by
\begin{equation}
\label{eq:changespace}
\begin{cases}
\mathbf x^{TB}=\displaystyle\mathbf x^{TX}\left(1-\frac{U}{c^2}-L\right)-\frac{1}{2}\left(\frac{\mathbf v_X^{TB}\cdot \mathbf x^{TX}}{c^2}\right)\mathbf v_X^{TB},\\
\mathbf v^{TB}=\displaystyle\left[\mathbf v^{TX}\left(1-\frac{U}{c^2}-L\right)-\frac{1}{2}\left(\frac{\mathbf v_X^{TB}\cdot \mathbf v^{TX}}{c^2}\right)\mathbf v_X^{TB}\right]\frac{dt^{TX}}{dt^{TB}}.
\end{cases}
\end{equation}
Because of the chain rule of derivative, for the velocities $\mathbf v^{TB}$ we have to account for the factor $dt^{TX}/dt^{TB}$ defined in Eq.~\eqref{eq:changetime}.

In Orbit14, in order to fully account for the relativistic corrections to the radio-science experiments, we implemented the multi-chart approach. This choice is also motivated by the fact that Eq.~\eqref{gravexp}, which describes the perturbation due to the gravitational field of the primary body, is formally correct if used in the local chart of the body. \citet{GODARD-etal_2012} investigated the differences in the propagation of the spacecraft dynamics using the multi and single-chart approaches. They found that, when accounting for all the necessary relativistic transformations to pass from one formulation to the other, the differences in the position and velocity of the probe are very small.

\begin{figure}[t]
\centering
\includegraphics[scale=0.28]{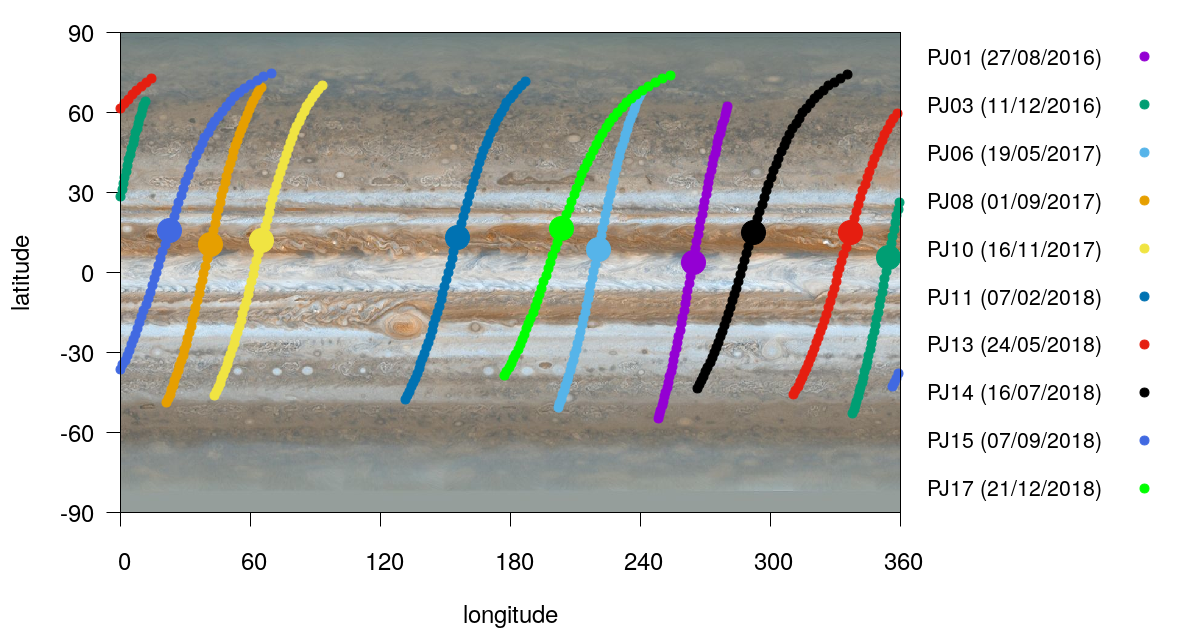}
\caption{Tracks of the Juno spacecraft over Jupiter's surface, represented in the so-called System III coordinates. As the cyclones present in the atmosphere of Jupiter, like the Great Red Spot, move in longitude, their position in the plot is not indicative of their actual location.}
\label{fig:gtrack}
\end{figure}

\section{Juno gravity experiment}
\label{sec:juno}
Juno is a NASA space mission orbiting around Jupiter since 2016 \citep{BOLTON-etal_2017}. Its orbit is polar and highly elliptic, so that, for most of time, the spacecraft is far from the planet. However, at perijoves, Juno reaches altitudes over planet's surface of a few thousands of kilometers, so that Doppler data are very sensitive to the gravitational field of Jupiter, making Juno an extraordinary source of information to constrain the interior of the planet. Although the original plan provided an engine burn to pass from an initial orbital period of $53.4$ days to just $14$, a malfunction forced the Science Team to cancel it. Therefore, the nominal mission accounts for a total of 34 perijoves and its end is set to July 2021. At the time of writing of this paper (March 2021), Juno completed 32 out of 34 perijoves.

In this section, we present a brief summary of the current knowledge achieved so far thanks to the Juno gravity experiment and we show the experiments we performed with the Orbit14 software using radio-science data from the first half of the nominal mission.

\subsection{Current state of the mission}
\label{subsec:juno_cur}
In the years before Juno data were available, several studies have been published about the modeling of the internal processes of Jupiter. Scientists implemented sophisticated equations of states to describe the interior of the planet \citep{HELLED-etal_2011,NETTELMANN-etal_2012,MIGUEL-etal_2016,MILITZER-etal_2016} and to evaluate their effects on its gravitational field. Moreover, they investigated the signatures due to the depth of jet streams \citep{GALANTI-KASPI_2017}, differential rotation \citep{KASPI-etal_2017}, tidal response \citep{WAHL-etal_2016} and normal modes \citep{DURANTE-etal_2017}. These models were meant as a solid basis for the interpretation of the Juno gravity data.

Studies were carried out also to asses the performances of the Juno gravity experiment. Simulations preceding the arrival of the probe in the Jovian system investigated the final uncertainty with which the gravitational field \citep{FINOCCHIARO-IESS_2010} and other parameters, like Jupiter's polar moment of inertia \citep{LEMAISTRE-etal_2016}, would have been determined. The CMG of Pisa performed simulations with the Orbit14 software, carrying out covariance analyses of the gravitational and tidal parameters, the direction of the pole and the angular momentum of Jupiter \citep{SERRA-etal_2016}. They investigated also the possibility to use ring-shaped mascons to model the local gravity field at the latitudes covered by Juno's perijoves \citep{SERRA-etal_2019b}.

\begin{figure}[t]
\centering
\includegraphics[scale=0.3]{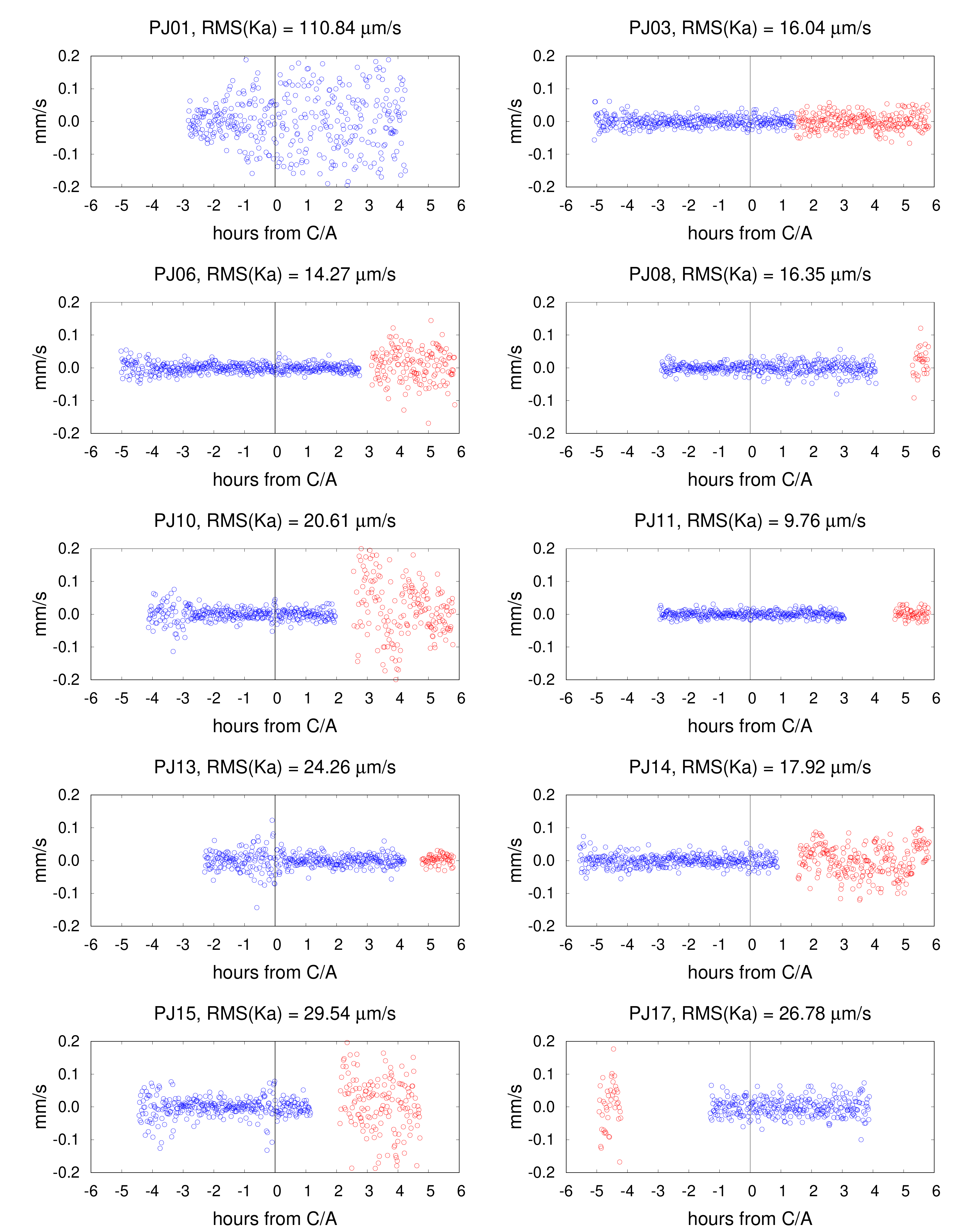}
\caption{Residuals of the ten arcs considered in the Juno gravity experiment. In blue, we plotted the Ka-band data collected at the close approach, while, in red, the X-band data obtained with the support of secondary tracking stations. For all plots, we considered the same ranges of the x and y axis, in order to have a clear comparison between the arcs, and we reported the RMS for the Ka-band data. Only for PJ01 the Ka-band residuals spread over the $[-0.2,0.2]$ mm/s interval.}
\label{fig:pj17res}
\end{figure}

Through the data of the first two perijoves, PJ01 and PJ02, it was possible to have a first great improvement on the knowledge of the gravitational coefficients of Jupiter \citep{FOLKNER-etal_2017}. However, only when the gravity perijoves PJ03 and PJ06 were performed, the north-south asymmetry of the gravitational field was revealed \citep{IESS-etal_2018,SERRA-etal_2019}. The significant improvement of the even zonal harmonics and the detection of the odd zonal harmonics provided strong constraints on the interior of Jupiter. In particular, from the analysis of the even harmonics, \citet{GUILLOT-etal_2018} found that, although Jupiter's atmosphere shows zones with very different rotation speeds, the deep interior rotates nearly as a rigid body. Moreover, from the odd harmonics, \citet{KASPI-etal_2018} found that the observed jet streams persist up to a depth of about 3,000 kilometers from the cloud tops. The computed zonal harmonics provided constraints also on the density profile of Jupiter, suggesting the presence of an extended dilute core \citep{WAHL-etal_2017}. A comprehensive review of Jupiter's interior as revealed so far by means of Juno's measurements of the gravitational and magnetic fields is presented in \citet{STEVENSON_2020}.

A further analysis of the gravity experiment was carried out by \citet{DURANTE-etal_2020} considering Doppler data up to PJ17, which marked the middle of the nominal mission. Apart from improving the uncertainty of the gravitational parameters, they obtained first hints of a dynamical response of the tides of Jupiter raised by the Galilean satellites. They found also small signals in the Doppler data at the close approaches, which could be related to longitudinal or time-varying features of the gravitational field of Jupiter. In the next section, we use the same setup of \citet{DURANTE-etal_2020} and we show the results of OD experiments with Orbit14. We want to stress that most of these results have been already presented in \citet{DURANTE-etal_2020} and that here we focus on the ability of Orbit14 to deal with a real and complex radio-science experiment.

\begin{figure}[t]
\includegraphics[scale=0.42]{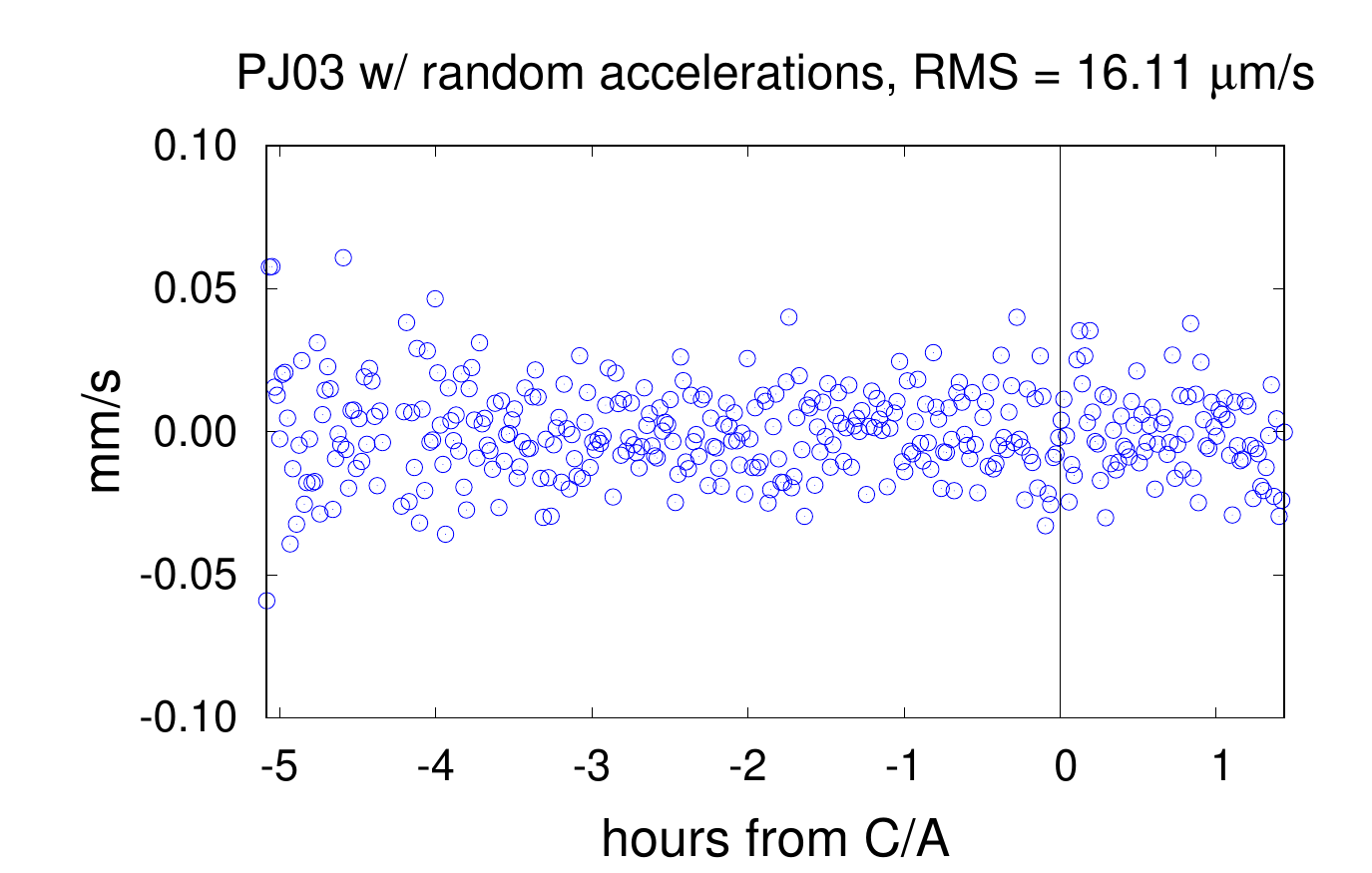}
\includegraphics[scale=0.42]{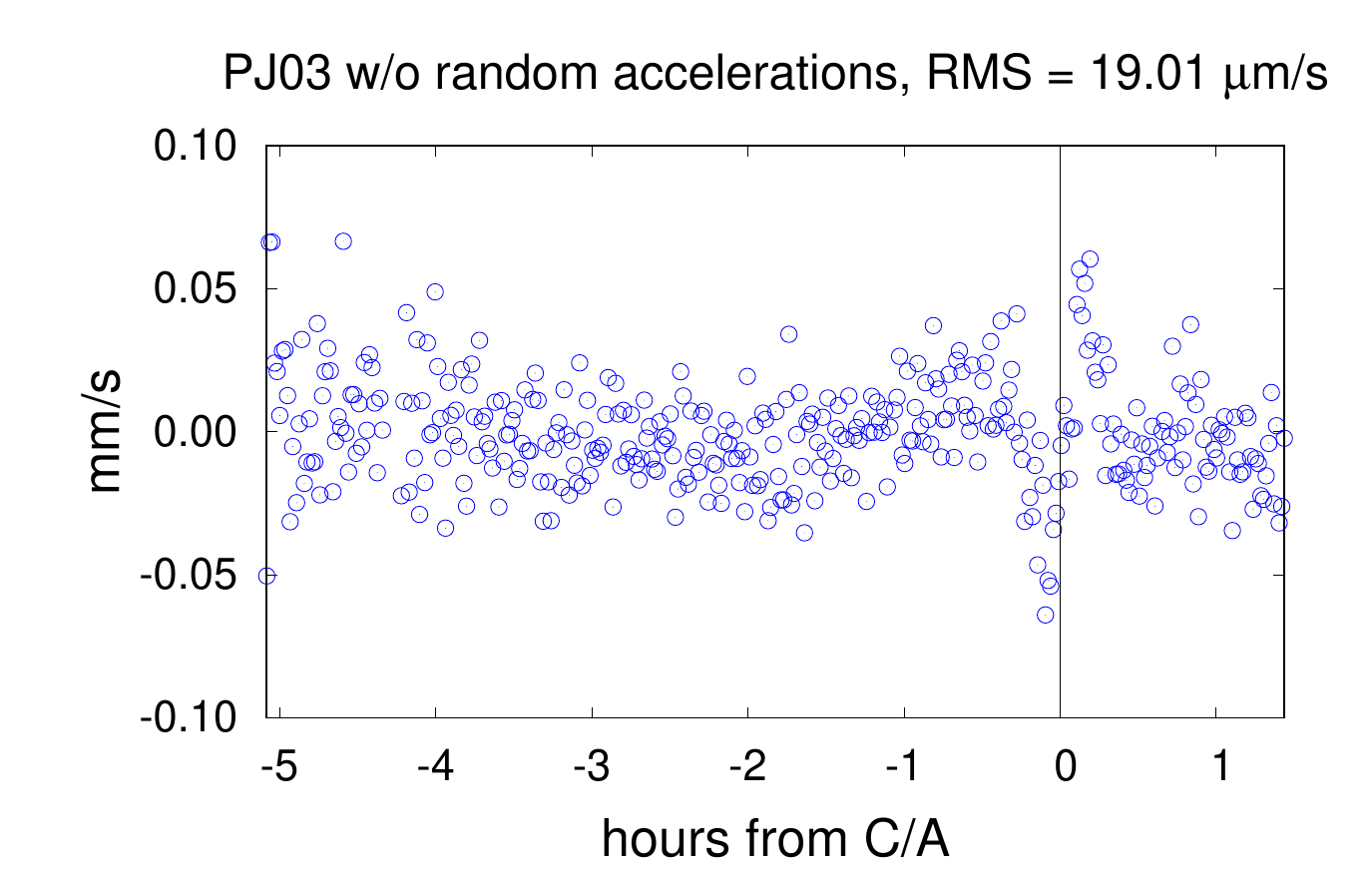}
\caption{Residuals of PJ03 (multi-arc experiment) in the case we estimate random accelerations (left) and in the case we do not (right). In the plot on the right, small signals at the close approach are not absorbed.}
\label{fig:resrandom}
\end{figure}

\subsection{Experiment with Orbit14}
\label{subsec:juno_exp}
During the first half of the nominal mission, 10 perijoves out of 17 were dedicated to the gravity experiment. Juno's ground tracks for these perijoves are shown in \figurename~\ref{fig:gtrack}, where we can appreciate their diagonal path due to the fast rotation of the planet. From the same figure, we can note that the points of closest approach (C/A) over Jupiter's surface take place at latitudes between $4^\circ$ and $17^\circ$. For gravity-dedicated perijoves, the high-gain antenna of the spacecraft is pointed toward the Earth and provides the Ka-band link with the ground-based stations. Apart from PJ01 and PJ13, during the close approaches, the spacecraft was tracked by the Deep Space Station 25 (DSS-25) at Goldstone and the gravity experiment was supported by both Ka/Ka and X/X links. Moreover, for almost all the ten perijoves, close approaches data were preceded or followed by a few hours of X/X data collected from other tracking stations of the Deep Space Network (DSN).

For each PJ, we consider the Doppler data collected at the close approach and also the observations from supporting stations, provided that between these two subsequent sets of data Juno did not perform any maneuver. The weights we assign to the observables are obtained from an aposteriori evaluation of the RMS of their residuals, obtained with a single-arc fit. For fitting the data of all ten PJs, we use a pure multi-arc strategy, where each arc corresponds to a perijove. Finally, for propagating the dynamics of the spacecraft and compute the observables, we take the states of the planets (including Earth and Jupiter) from JPL ephemerides DE438 and the states of the Jovian satellites from JPL ephemerides JUP310\footnote{https://naif.jpl.nasa.gov/pub/naif/JUNO/kernels/spk/}. Because of the limited time span we are considering in the experiment, we can use a simplified model for the precession of Jupiter's spin:
\begin{equation}
\label{eq:linspin}
\begin{cases}
\alpha(t)=\alpha_0+\dot \alpha (t-t_0),\\
\delta(t)=\delta_0+\dot \delta (t-t_0).
\end{cases}
\end{equation}
The linear model in Eq.~\eqref{eq:linspin} depends on the direction of the pole $(\alpha_0,\delta_0)$ at a certain initial time (e.g, J2000 epoch) and on its rate $(\dot \alpha,\dot \delta)$, which we assume to be constant. Although this is a rough approximation with respect to the model presented in \citet{ARCHINAL-etal_2018}, it is sufficient to obtain a good fit of the data (see \citealt{DURANTE-etal_2020}).

The main goal of the experiment is to estimate the gravity field of Jupiter along with its tidal parameters. Moreover, in order to absorb all the signals in the data and flatten the residuals, we need to solve for several parameters:
\begin{itemize}
\setlength\itemsep{0.2em}
\item gravitational parameter of Jupiter $Gm_0$ (1),
\item zonal gravity coefficients from $J_2$ to $J_{30}$ (29),
\item tesseral gravity coefficients of degree $2$ (4),
\item tidal parameters $k_{22},\ k_{31},\ k_{33},\ k_{42}$ and $k_{44}$ (5),
\item pole direction and (constant) rate (4),
\item spacecraft's state (local, 6 per arc),
\item SRP coefficient (local, 1 per arc),
\item random accelerations coefficients (local, 36 per arc);
\end{itemize}
in the parentheses, we reported the total number of the parameters.

Random accelerations are empirical accelerations used to absorb tiny signals near C/A. For the Juno gravity experiment, they are necessary when we process a large number of perijoves. Indeed, as it has been observed by \citet{DURANTE-etal_2020}, a purely zonal field is not able to fit properly the data at the closest approaches (see \figurename~\ref{fig:resrandom}). Random accelerations are modeled as piece-wise constant functions: following \citet{DURANTE-etal_2020}, we add them to the dynamical equations of the spacecraft only in the time span $[-1,+1]$ hours from C/A. We subdivide it in intervals of $10$ minutes for which we estimate the constant value of the empirical acceleration. At this stage of the mission, we use this formulation because the physical process responsible for the signals in the Doppler data is currently under investigation, even though it is probably related to unmodeled effects of the gravitational field of Jupiter.

For almost all the parameters we solve for, we use large apriori, as Juno data alone allow to improve their knowledge with unprecedented accuracy. Only for the gravitational parameter of Jupiter $Gm_0$ we use strict apriori, as the Galileo mission \citep{JACOBSON-etal_2000} provides a better estimation than the one we obtain with Juno using only perijoves data. Recently, \citet{NOTARO-etal_2021} showed that including Juno data far from the perijoves allows to improve the determination of Jupiter $Gm_0$. Also for random accelerations we use significant apriori ($\sigma^P=2\times 10^{-6}$ cm/s$^2$), as we want to limit their magnitudes to fit the observed remaining signals at perijoves (\figurename~\ref{fig:resrandom}). For building the first guess of the fit parameters vector, we rely on values presented in \citet{IESS-etal_2018} (gravitational field), \citet{WAHL-etal_2016} (tides) and \citet{ARCHINAL-etal_2018} (pole). First guess for the initial conditions of the spacecraft are taken from SPICE kernels provided by NASA.

We use Orbit14 software to process the data. More precisely, we consider Doppler data only and we compress them to an integration time of $60$ s in order to remove very short-time signals not related to gravity effects. The OD procedure converges very quickly: at the third iteration the norm of the variation of the parameters is already very small.

In \figurename~\ref{fig:pj17res}, we plotted the residuals of all the ten arcs we are considering in the OD experiment. Residuals at close approaches (blue dots) appear to be flat, apart from some groups of them that suffer a temporary increase of the noise. Their Root Mean Square (RMS), reported on the top of the plots, can be very different from one arc to another, providing information about the quality of the data. However, apart from PJ01, which shows a very high noise due to the effect of the interplanetary plasma on the X-band uplink, the RMS of the other arcs is satisfying. In particular, for a few of them we find an RMS of about $0.015$ mm/s or smaller, which is very near the data nominal accuracy of $0.012$ mm/s. The other residuals (red dots) are computed from data collected by supporting stations. As for these data only the X/X link was available, they suffer the lack of plasma calibrations, showing a larger spread distribution of the residuals, which are not perfectly flat. Despite their inferior quality with respect to the observations from the main station, these data contribute to decorrelate the initial conditions of the spacecraft and the dynamical parameters of Jupiter. Therefore, their inclusion in the gravity experiment is recommended (see \citealt{SERRA-etal_2019}).

As we already mentioned, random accelerations are essential to flatten the residuals near C/A. In \figurename~\ref{fig:resrandom}, we compared the residuals of PJ03 in the case we estimate random accelerations (as in our nominal setup) and in the case we do not. In the second case, a clear signal remains in the post-fit residuals and increases significantly their RMS.

\begin{figure}[t]
\centering
\includegraphics[scale=0.6]{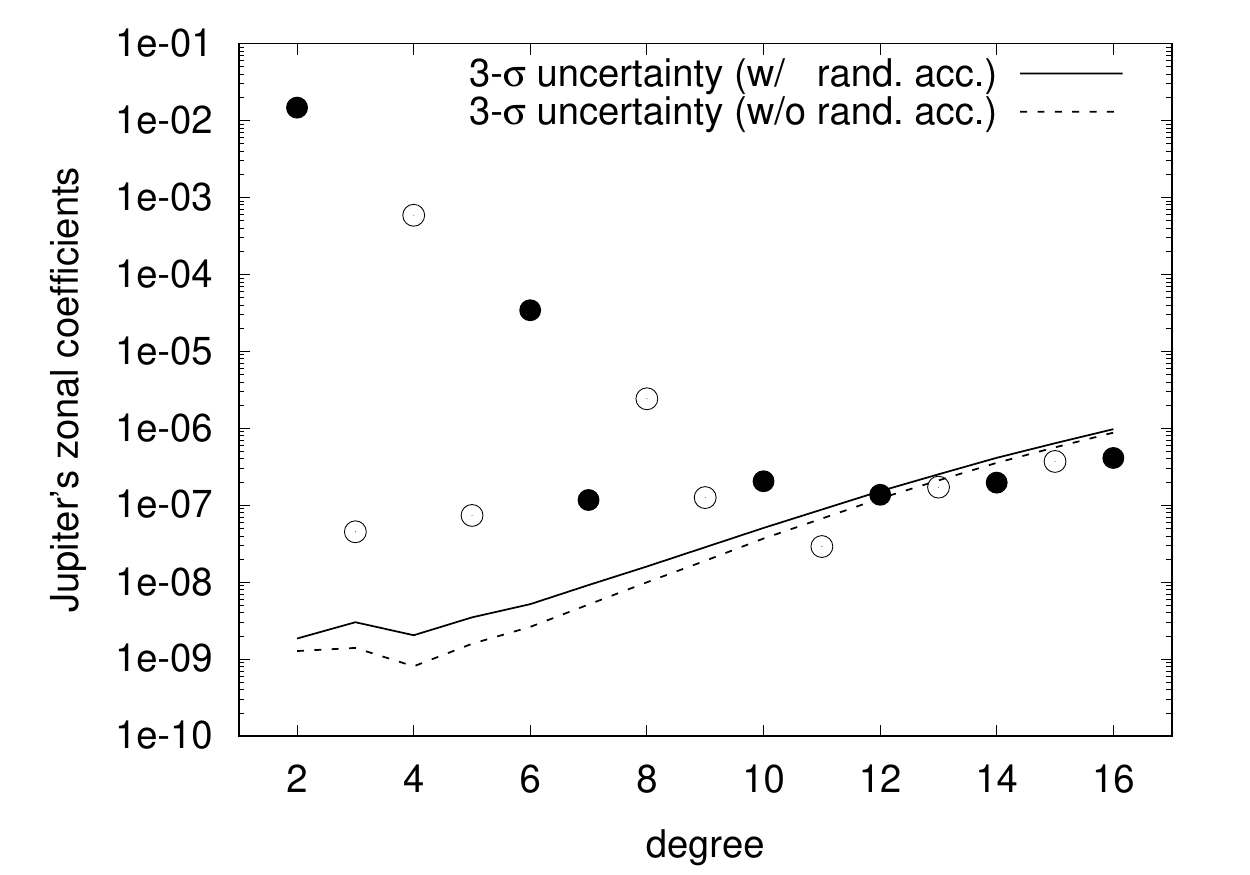}
\caption{Estimated values of Jupiter's zonal coefficients. Filled circles indicate negative values, while empty circles indicate positive values. From the black solid line, which represents the formal uncertainty of the parameters, we see that only zonal coefficients $J_{\ell}$ up to $\ell=10$ are clearly determined. The dashed line is the formal uncertainty of the zonal parameters in the case we do not solve for random accelerations.}
\label{fig:zonalvalue}
\end{figure}

\begin{figure}[t]
\centering
\includegraphics[scale=0.75]{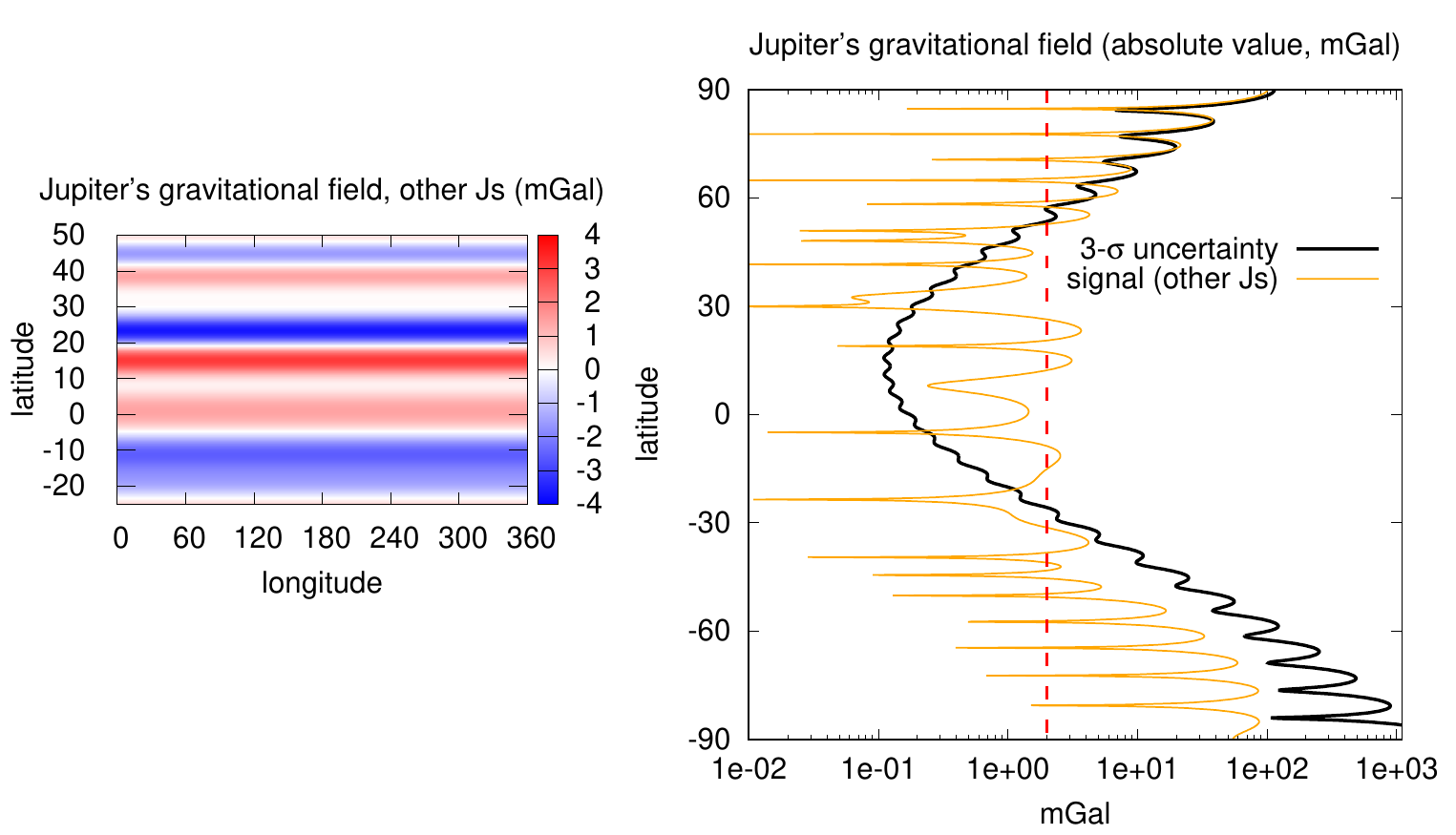}
\caption{On the left, Jupiter's gravitational field removing the contribution of the main harmonics ($J_2$, $J_4$, $J_6$ and $J_8$). We reported the field only in the latitude interval $[-25^\circ,50^\circ]$, as its formal uncertainty grows up toward the polar regions and the signal-to-noise becomes smaller than $1$ outside the considered interval. This is described by the plot on the right, where the black line is the $3$-$\sigma$ uncertainty of the gravitational field and the orange line is the signal due to the minor zonal harmonics. The red vertical line is set to $2$ mGal.}
\label{fig:geofieldother}
\end{figure}

In light of the good fitting of the data, we can now present the results we obtained about the determination of the parameters. In \figurename~\ref{fig:zonalvalue}, we reported the estimated value and uncertainty of the zonal gravitational coefficients $J_\ell$ of Jupiter. These parameters are clearly determined up to degree $10$, while, for higher degrees, the formal uncertainty, which increases with $\ell$, is greater than the estimated values. The even harmonics are clearly dominant with respect to the odd ones, due to the strong symmetry with respect to the equator, common to gas giants. However, this symmetry is not perfect, as the odd zonal harmonics are different from $0$ \citep{IESS-etal_2018}. In \figurename~\ref{fig:geofieldother}, we plotted the gravitational field over Jupiter's surface removing the main harmonics $J_2$, $J_4$, $J_6$ and $J_8$: the asymmetry with respect to the equator is clear from the different colors present in the south and north hemispheres. As represented in the same figure, the uncertainty of the gravitational field grows quickly toward the poles, because of their poor coverage by Juno. Moreover, the field has not the same resolution with respect to the equator, as the latitude of the closest approaches is slightly shifted toward the north pole of the planet (see \figurename~\ref{fig:gtrack}).

Although zonal harmonics approximate with great accuracy the gravitational field of Jupiter, they are not sufficient to have a perfect fit of the data. So far we found that the estimated value of the tesseral harmonics are statistically null (see \tablename~\ref{tab:tesseral}), however, the fact that non-negligible signals remain in the Doppler data suggests that something lacks in the model. Random accelerations are an ad-hoc expedient to easily absorb signals at the closest approach, but they are not related to any physical effect. The signals we observe can be due to local longitudinal or time-varying features of the gravitational field of Jupiter, such as normal modes in the interior of the planet (see \citealt{DURANTE-etal_2017}). It is worth noting that the estimation of random accelerations have significant consequences in the OD, as they make the final formal uncertainty of the gravitational field increase (see \figurename~\ref{fig:zonalvalue}).

\begin{table}
{\small
\begin{center}
\noindent\begin{tabular}{l|c|cc}
\toprule
 & axial symmetry & estimated value & $3$-$\sigma$ uncertainty
 \\
\midrule
$C_{21}$  & $0$ & $\phantom{0}2.3\times 10^{-10}$ & $19.5\times 10^{-10}$ \\
$S_{21}$  & $0$ & $\phantom{0}8.2\times 10^{-10}$ & $15.9\times 10^{-10}$ \\
$C_{22}$  & $0$ & $\phantom{0}0.7\times 10^{-10}$ & $\phantom{0}9.8\times 10^{-10}$ \\
$S_{22}$  & $0$ & $\phantom{0}5.0\times 10^{-10}$ & $\phantom{0}9.8\times 10^{-10}$ \\
\bottomrule
  \end{tabular}
\end{center}
}
\caption{Estimated values and uncertainties of Jupiter's tesseral coefficients of degree $2$. All of them are statistically compatible with $0$.}
\label{tab:tesseral}
{\small
\begin{center}
\noindent\begin{tabular}{l|c|cc}
\toprule
 & \citet{WAHL-etal_2016} & estimated value & $3$-$\sigma$ uncertainty
 \\
\midrule
$k_{22}$  & $0.590$ & $0.559$ & $0.017$ \\
$k_{31}$  & $0.194$ & $0.225$ & $0.045$ \\
$k_{33}$  & $0.244$ & $0.297$ & $0.113$ \\
$k_{42}$  & $1.787$ & $1.286$ & $0.175$ \\
$k_{44}$  & $0.139$ & $0.166$ & $0.397$ \\
\bottomrule
  \end{tabular}
\end{center}
}
\caption{Estimated values and uncertainties of Jupiter's tidal parameters. In the second column, we reported the values presented in \citet{WAHL-etal_2016} which describe a static tidal response of Jupiter due to Io (response due to other satellites is almost equal, apart from $k_{42}$).}
\label{tab:tides}
\end{table}

Finally, results on tidal parameters up to degree $4$ are presented in \tablename~\ref{tab:tides}. In the same table, we reported their theoretical values computed by \citet{WAHL-etal_2016} assuming a static response of the tides raised by the Galilean satellites. We note that not all tidal parameters are estimated, as for some of them the theoretical value is null and small deviations would be hardly revealed. As already found by \cite{DURANTE-etal_2020}, the estimated values of $k_{22}$ and $k_{42}$ are significantly different from their static values, giving a first hint of possible dynamical contribution due to interaction of the tidal perturbations with the interior of Jupiter \citep{WAHL-etal_2020}. It is worth noting that, in the experiment, we used tidal parameters common to all the Galilean moons, not taking into account possible differences in the tidal response between the satellites. A more accurate model must include satellite-dependent parameters \citep{NOTARO-etal_2019}. However, as showed in \citet{DURANTE-etal_2020}, with this formulation the formal uncertainty of the tidal parameters degrades greatly.

In conclusion, we presented an analysis of the Juno radio-science data up to PJ17 carried out with the Orbit14 software. However, Juno performed several other perijoves and many others are planned in the extended mission. For the future experiments with a larger amount of data, some aspects need to be further investigated. First of all, it will be important to see if the remaining signals at the closest approaches will increase in magnitude and if random accelerations will be able to still absorb them or other models related to physical processes will be required (see \citealt{DURANTE-etal_2017,PARISI-etal_2020}). Moreover, the use of a satellite-dependent formulation of the tides on Jupiter will be essential to confirm possible deviations of the tidal parameters of Jupiter from their static values (see \citealt{NOTARO-etal_2019}). Finally, as the time span covered by the mission increases, a linear evolution of the pole of Jupiter is not adequate anymore, but it will be necessary to integrate a complete model of the orientation of Jupiter and to include its polar moment of inertia to the list of the fit parameters (see \citealt{LEMAISTRE-etal_2016}). Orbit14 has been recently improved to deal with all these aspects of the future investigation and an update of the Juno gravity experiment is expected to be published by the Juno team in due course.

\section{BepiColombo MORE}
\label{sec:bc}
BepiColombo is a joint ESA and JAXA mission for the exploration of Mercury \citep{BENKHOFF-etal_2010}. The mission was launched in October 2018 and it is planned for orbit insertion around the planet in December 2025. It is composed by two spacecrafts, the Mercury Planetary Orbiter (MPO) and the Mercury Magnetospheric Orbiter (MMO), mounted on a common transfer module. At Mercury arrival, they will separate and move to two different orbits around the planet: the MPO altitude range is expected to vary between $400$ and $1500$ km, the MMO altitude ranges between $590$ and $11639$ km. The Mercury Orbiter Radio science Experiment (MORE) is one of the experiments on-board the MPO spacecraft, devised to enable a better understanding of the geodesy and geophysics of Mercury, on one side, and of fundamental physics, on the other (see, e.g., \citealt{MILANI-etal_2001,MILANI-etal_2002,IESS-etal_2009}). Thanks to full on-board and on-ground instrumentation capable to perform very precise tracking from the Earth \citep{IESS-BOSCAGLI_2001}, MORE will have the chance to determine with a very high accuracy the Mercury-centric orbit of the spacecraft and the heliocentric orbit of Mercury and the Earth. This will enable, in turn, to determine the gravitational field (gravimetry experiment) and the rotational state (rotation experiment) of the planet. Moreover, taking advantage from the fact that Mercury is the nearest planet to the Sun, MORE will allow to perform an accurate test of relativistic theories of gravitation, constraining simultaneously the value of some parameters of interest (relativity experiment).

\begin{table}
{\small
\begin{center}
    \begin{tabular}{|l|l|}
    \hline
     General   & State vectors of Mercury and the Earth \\
               & State vector of MPO \\
               & Accelerometer calibrations \\
               & Dump maneuvers calibrations \\
        \hline
        Relativity & PN parameters ($\beta$, $\gamma$, $\eta$, $\alpha_1$, $\alpha_2$)\\
                              & Solar parameters ($J_{2\odot}$, $\mu_{\odot}$, $\zeta$, $GS_\odot$)\\
         \hline
         Gravimetry & Spherical harmonics coefficients ($C_{\ell m}$ and $S_{\ell m}$)\\
           & Love number ($k_2$)\\
          \hline
          Rotation & Pole direction ($\alpha$, $\delta$) \\
                              & Amplitudes of the libration ($\epsilon_1$, $\epsilon_2$) \\
           \hline
    \end{tabular}
\end{center}
}
    \caption{List of the solve-for parameters for BepiColombo divided according to the corresponding experiment.}
    \label{tabBC}
\end{table}

\subsection{Simulation setup and nominal results}
\label{subsec:bc_sim}
The main goal of the gravimetry experiment is to characterize the gravitational field of Mercury at medium and short scales (up to about a few hundreds of kilometers). Thanks to the proximity of the MPO spacecraft to planet's surface and the thousands of orbital revolutions that it will perform around Mercury, it will be possible to determine gravitational coefficients up to degree and order $25$ with a signal-to-noise ratio around $10$ \citep{CICALO-etal_2016,IMPERI-etal_2018}. Moreover, the rotation experiment will allow to estimate accurately the direction of the pole of the planet and the main amplitudes of its libration in longitude \citep{PFYFFER-etal_2011,SCHETTINO-etal_2017}, whose periods are 88 days (forced term by Mercury's rotation) and 11.86 years (resonant term with Jupiter) \citep{PEALE-etal_2009}. Both the experiments will be an incredible source of information to constrain the interior of Mercury \citep{SPOHN-etal_2001}.

The relativity experiment will investigate possible deviations from the General Relativity theory, thanks to the excellent determination of the orbit of Mercury provided by MORE. In particular, the main goal is to estimate some of the post-Newtonian parameters and verify if their values are in accordance with the ones expected by the General Relativity or not (see \tablename~\ref{tab:pntab}). In order to have a reliable determination, it is necessary to estimate other parameters that affect the relativity experiment, such as the solar oblateness factor $J_{2\odot}$, the gravitational mass of the Sun $\mu_\odot=GM_\odot$ and its time derivative $\zeta$, and the solar angular momentum $GS_\odot$ (see, e.g, \citealt{GENOVA-etal_2018}).

BepiColombo mission also counts the Italian Spring Accelerometer (ISA) \citep{IAFOLLA-NOZZOLI_2001}, which is mounted on the MPO spacecraft. ISA will provide accurate measurements of the strong non-gravitational forces acting on the probe. Moreover, the orbit design of the mission comprehends also frequent dump maneuvers to maintain the desired attitude of the spacecraft. In terms of parameters estimation, it is then necessary to add calibration coefficients both for the accelerometer and the maneuvers into the fit. %Furthermore, systematic terms should be considered in the error model of the radio-science data, especially for the range measurements (see \citealt{SCHETTINO-etal_2016b} for details). However, they are unimportant for the purposes of a covariance analysis, as presented in Sect.~\ref{sec:od}, thus we will neglect it in the following.

\begin{figure}[t]
\centering
\includegraphics[scale=0.7]{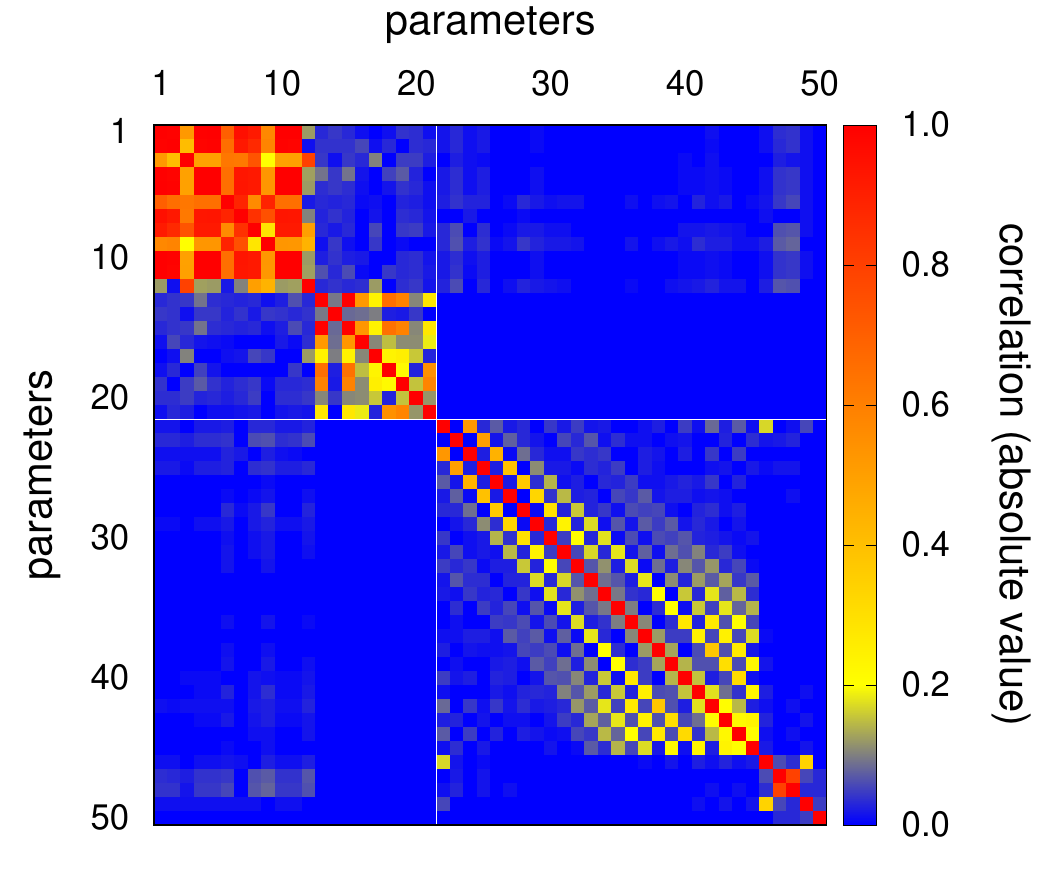}
\caption{Representation of the correlations between the global parameters through a color map. Because the number of gravitational coefficients is huge, we reported the correlations of the zonal coefficients only. The parameters are, in order, initial conditions of Earth and Mercury (1-12), PN and solar parameters (13-21), zonal gravitational coefficients (22-45), Love number $k_2$ (46) and rotation parameters (47-50). The white lines highlight the separation between relativity parameters and gravity-rotation parameters.}
\label{fig:tabcorr}
\end{figure}

For the simulations of the BepiColombo radio-science experiment, we assume that the radio tracking observables are affected only by random effects, with a nominal standard deviation of $\sigma_r = 30$ cm for the 2-way range at $300$ s and $\sigma_{rr} = 3 \times 10^{-4}$ cm/s for the 2-way range-rate at $1000$ s, as provided by the KaT. The end-to-end simulations have a nominal duration of $1$ year, subdivided into arcs of about $1$ day, and start on 14 March 2026, which is the day the MPO spacecraft reaches its science orbit. Further details concerning the gravimetry and rotation experiments can be found in \citet{CICALO-etal_2016}, \citet{SCHETTINO-TOMMEI_2016} and \citet{SCHETTINO-etal_2017}, while an extensive analysis of the the relativity experiment is presented in \citet{SCHETTINO-TOMMEI_2016} and \citet{SCHETTINO-etal_2018}. Moreover, a comprehensive review of the MORE experiment has been recently published in \citet{IESS-etal_2021}.

In the end, the list of the parameters of interest for the three experiments of MORE (relativity, gravimetry, rotation) are the following:
\begin{itemize}
\setlength\itemsep{0.2em}
    \item state vectors of Mercury and the Earth\footnote{Actually, we compute the state vector of the Earth-Moon barycenter (EMB).} (12),
    \item post-Newtonian parameters: $\beta$, $\gamma$, $\eta$, $\alpha_1$, $\alpha_2$ (5),
    \item solar parameters: $J_{2\odot}$, $\mu_\odot$, $\zeta$ and $GS_\odot$ (4),
    \item harmonic gravitational coefficients of Mercury up to degree and order 25 (672),
    \item Love number of Mercury $k_2$ (1),
    \item pole direction of Mercury (2),
    \item main amplitudes of the libration in longitude of Mercury (2),
    \item state vector of the MPO spacecraft (local, 6 per arc),
    \item calibration coefficients for the accelerometer (local, 6 per arc),
    \item calibration coefficients for the dump maneuvers (local, 3 per maneuver per arc);
\end{itemize}
in the parentheses, we reported the total number of the parameters. For the sake of clarity, the solve-for parameters are summarized in Table \ref{tabBC}.

All the parameters above can be determined simultaneously within a global least squares fit (see Sect.~\ref{sec:od}), adopting a constrained multiarc strategy (see Sect.~\ref{subsec:adv_mult}). The complete process of simulation is performed in two steps. The first step consists in simulating the radio observables and the preliminary orbit of the spacecraft. In the second step, the simulated data are processed by the differential corrector, which performs the least squares fit (see \citealt{SCHETTINO-TOMMEI_2016} for further details). During the differential correction process, we make use of some assumptions. First of all, following the strategy described in Sect.~\ref{subsec:adv_ap}, we set an apriori constraint on each solve-for parameter, given by the present knowledge of the parameter. In the case of the PN parameter $\gamma$, we set a more tight constraint\footnote{At present, the uncertainty by which the $\gamma$ parameter is equal to unity has been provided by the Cassini mission at $2.5\times 10^{-5}$ level \citep{BERTOTTI-etal_2003}. The conservative limit we derived by simulations in \citet{SERRA-etal_2018} can be set at $1\times 10^{-5}$ level, which is adopted as the apriori constraint for the in-orbit simulations.}, as the result of a set of simulations performed by the authors on the Superior Conjunction Experiment (SCE), which will be held during the cruise phase of BepiColombo from 2021 to 2024 (see \citealt{SERRA-etal_2018}). Moreover, we adopt  the following important assumption: we link all the PN parameters by the Nordtvedt equation \citep{NORDTVEDT_1970}
\begin{equation}
\label{eq:eta}
\eta = 4(\beta -1) - (\gamma -1) -\alpha_1 -\frac{2}{3}\alpha_2\,,
\end{equation}
which means that we are considering only metric theories of gravity. This constraint has been adopted since the initial conception of the relativity experiment and is motivated by the fact that $\beta$ and $J_{2\odot}$ are expected to show an almost 100\% correlation, since their dynamical effect on the orbit of Mercury is comparable (see, e.g., \citealt{MILANI-etal_2002} for an extensive discussion of this symmetry in the case of the MORE relativity experiment). The addition of the constraint removes the degeneracy between $\beta$ and $J_{2\odot}$, but this result is, in turn, obtained at the cost of forcing an almost 100\% correlation between $\beta$ and $\eta$.

The results of the MORE simulations have been already extensively described in a number of recent publications (see \citealt{SCHETTINO-TOMMEI_2016,SCHETTINO-etal_2018,IMPERI-etal_2018,DEMARCHI-CASCIOLI_2020}). An interesting and remarkable point that we would like to stress here concerns the fact that, although the MORE experiment has been conceived and devised as a unique experiment, in practice, the relativity and the gravimetry-rotation experiments can be considered and performed as two independent experiments. As a matter of fact, the results of the two experiments turn out to be almost perfectly orthogonal. As a consequence, performing a global least squares fit or two separated fits with the same data does not change the results in terms of the covariance analysis.

This fact can be easily understood by looking at the correlations between pairs of parameters $x_i$ and $x_j$. Recalling the definition of correlation given in Eq.~\eqref{eq:formunc}, the two boundary cases can be interpreted in the following way: if $\mathrm{corr}(x_i,x_j)=1$, the two parameters cannot be simultaneously determined within the same least squares fit, since they produce the same dynamical effect and the available data are not capable to discriminate the effect due to one parameter from that due to the other; on the contrary, if $\mathrm{corr}(x_i,x_j)=0$, the two parameters turn out to be totally independent one from the other, thus including one or both in the fit does not undermine the determination of either parameter. In the case of MORE, what happens is that the estimated parameters arrange along blocks: each parameter is correlated with the parameters within the same block, while it shows an almost null correlation with the parameters of the other blocks. In particular, the relativity parameters correlate only marginally with the gravimetry-rotation parameters: this fact allows to handle them as two separate experiments. A general plot of the correlations between the solve-for parameters can be seen in \figurename~\ref{fig:tabcorr}: the relativity and gravimetry-rotation blocks can be clearly identified.

To elucidate this behavior, we can make two remarks. First, gravimetry involves short period phenomena, mainly related with the orbital motion of the MPO around Mercury, whose period is about 2.3 hours. On the other side, relativistic phenomena take place over much longer time scales, of the order of months or years, thus they can be inferred mainly by studying the heliocentric motion of the Earth and Mercury. Secondly, if we compare the standard deviation of range and range-rate over the same integration time, according to Gaussian statistics, we find that $\sigma_r/\sigma_{rr}\sim 10^5$ s. As a consequence, range data are more accurate than range-rate when observing phenomena taking place over time scales longer than some days (as the phenomena of interest for the relativity experiment), while the opposite holds in the case of gravimetry.

\begin{table}[t]
\[
\begin{array}{ccccc}
\toprule
\textrm{Parameter} & \textrm{Current knowledge}&\textrm{Sim. 1} & \textrm{Sim.  2} & \textrm{Sim. 3} \\
\midrule
\beta & 1.8 \times 10^{-5} & 3.33 \times 10^{-5}& 4.36 \times 10^{-6} & 3.33 \times 10^{-5} \\
\gamma & 2.3 \times 10^{-5} & 1.32 \times 10^{-6}& 1.32 \times 10^{-6}& 1.32 \times 10^{-6}\\
\eta & 4.5\times 10^{-4} & 1.34 \times 10^{-4} & 1.66 \times 10^{-5}& 1.34 \times 10^{-4}\\
\alpha_1 & 6.0 \times 10^{-6} & 1.16 \times 10^{-6} & 9.00 \times 10^{-7}& 1.16 \times 10^{-6}\\
\alpha_2 & 3.5\times 10^{-5} & 1.88 \times 10^{-7} & 1.87 \times 10^{-7}& 1.87 \times 10^{-7} \\
\mu_{\odot} & 8.0 \times 10^{15} & 1.08 \times 10^{14} & 8.08 \times 10^{13} & 1.08 \times 10^{14} \\
J_{2\odot} & 2.2\times 10^{-9} & 3.42 \times 10^{-9} & \ 9.35 \times 10^{-10} & 3.42 \times 10^{-9}\\
\zeta & \ 4.3 \times 10^{-14} & \ 4.03 \times 10^{-14} & \ 3.69 \times 10^{-14} & \ 4.02 \times 10^{-14}\\
\bottomrule
\end{array}
\]
\caption{Comparison of the formal uncertainty of the PPN parameters $\beta, \gamma,\eta,\alpha_1,\alpha_2$ and related solar parameters $\mu_{\odot},J_{2\odot},\zeta$ in four cases: current knowledge (see \citealp{SCHETTINO-etal_2020} for the references), a simulation where all parameters where solved for and no constraints were used (Sim. 1), a simulation where all parameters except four of them were solved for (descoping) and no constraints were used (Sim. 2), a simulation where all parameters were solved for and some apriori constraints were used (Sim. 3; see \citealp{SCHETTINO-etal_2016} for the apriori constraints). The uncertainty of $\mu_{\odot}$ is in cm$^3$/s$^2$ and of $\zeta$ in y$^{-1}$.}
\label{tab:relpar}
\end{table}

\subsection{An example: tests of alternative theories of gravity}
\label{subsec:bc_rg}
The relativity experiment has been originally conceived as an opportunity to further confirm the validity of Einstein's theory of General Relativity at the level of the solar system \citep{MILANI-etal_2002}. The majority of the solar system tests of gravitation can be set in the context of the slow-motion, weak field limit \citep{WILL_2014}, usually known as the post-Newtonian (PN) approximation. In this limit, the space-time metric can be written as an expansion about the Minkowski metric in terms of dimensionless gravitational potentials. In particular, adopting the parameterized PN (PPN) formalism, it is possible to express each potential term in the metric by means of a given parameter, which measures a general property of the metric (see \tablename~\ref{tab:pntab}). The MORE relativity experiment originated from the idea of investigating the dependence of the equation of motion from the PN parameters. Indeed, by isolating the effects of each parameter on the motion, it is possible to constrain the parameters values within some accuracy threshold, testing the validity of relativistic predictions. The standard test consists in determining the Eddington parameters $\gamma$ and $\beta$ (both expected to be equal to 1 in General Relativity), the Nordtvedt parameter $\eta$ and the preferred-frame effects parameters $\alpha_1$ and $\alpha_2$ (expected to be all null in General Relativity).

While recently a number of criticalities has emerged, in particular concerning the estimate of $\beta$ and $\eta$ \citep{DEMARCHI-etal_2016,DEMARCHI-CASCIOLI_2020}, the unprecedented accuracy that will be achieved in determining the orbit of Mercury around the Sun unfolds the possibility to test and constrain parameters which do not appear in the standard relativistic theory. In particular, we studied the possibility of testing alternative theories of gravity including a detectable space-time torsion, which is expected to vanish in General Relativity. All the details have been extensively described in \citet{SCHETTINO-etal_2020}, following the theoretical analysis given by \citet{MARCH-etal_2011}.

The same kind of analysis can be potentially applied to different alternative theories of gravity. Indeed, the basic idea is to suitably express the relativistic effect we want to test in a PPN framework, defining one or more parameters to describe the effect on the metric in the same way of PN parameters. In the case of space-time torsion, we included 3 independent torsion parameters (called $t_1$, $t_2$, $t_3$) in the solve-for list, assuming a nominal first-guess value equal to zero (as in General Relativity). Then, we estimated the torsion parameters in the global least squares fit together with the other relativity parameters and we checked: (1) if the torsion parameters can be determined and at which level of accuracy; (2) if the torsion parameters show a significant correlation with the other relativity parameters. In the case of high correlation, a general worsening of the solution is expected: more precisely, a deterioration of the formal uncertainty of the correlated parameters. In general, the addition of a new parameter in the solve-for list must be evaluated by considering the trade-off between the advantage to test additional dynamical effects and the possible worsening of the solution, which needs to be restrained. With the same kind of approach, in the near future we plan to test if other alternative theories of gravitation could be constrained, at the solar system level and within a PPN framework, by means of the MORE relativity experiment.

\begin{figure}[t]
\centering
\includegraphics[scale=0.7]{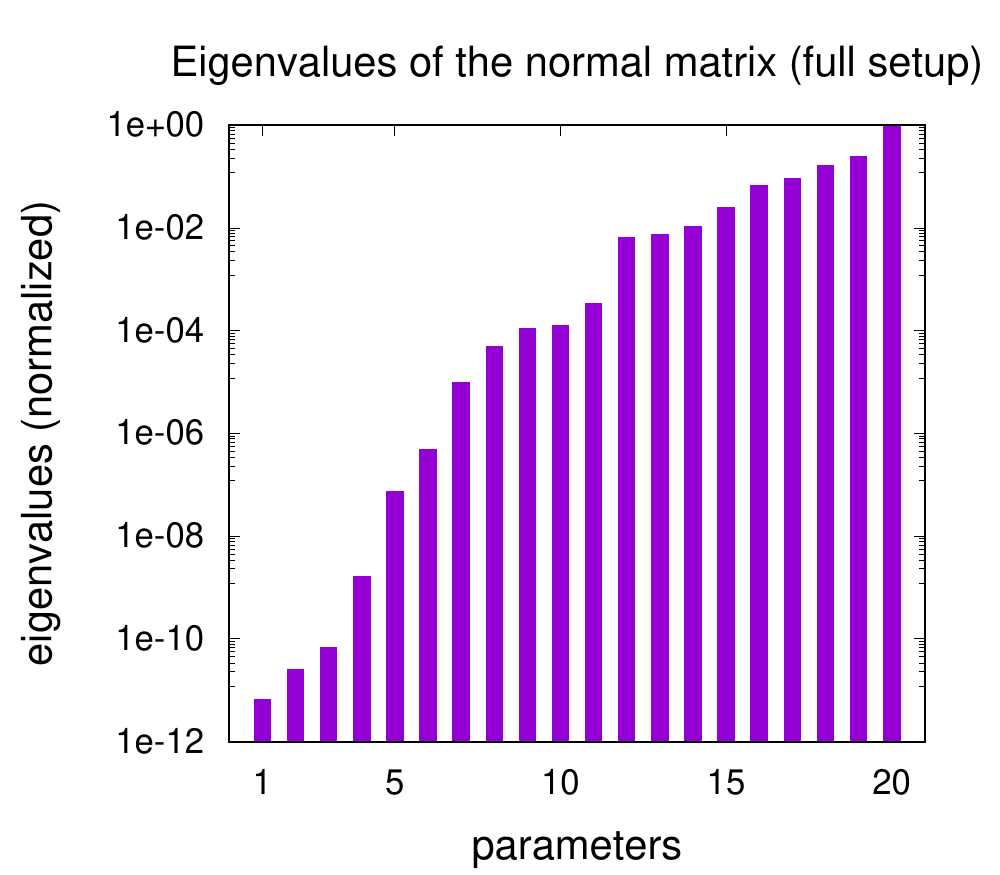}
\caption{Ordered eigenvalues of the global normal matrix in the case of a full MORE relativity experiment simulation. The global normal matrix has dimension 20x20, corresponding to solving for 6 initial conditions of EMB, 6 initial conditions of Mercury, 8 relativity and solar parameters.}
\label{fig:histfull}
\end{figure}

\subsection{Approximate rank deficiency in the MORE experiment}
\label{subsec:bc_rk}

Ever since the first simulations were performed, the MORE experiment has been believed to be affected by an approximate rank deficiency of order 4 \citep{MILANI-etal_2002,SCHETTINO-etal_2016}. This means that the normal matrix of the fit $C$ is ill-conditioned and has, at least, four eigenvalues close to zero, resulting in a very degraded solution (see Sect.~\ref{subsec:adv_rk}). As a matter of fact, the formal uncertainty of the PPN parameters, which we discussed in the previous section, are smaller, even by orders of magnitude, when we eliminate four parameters from the fit (descoping). This seems to show that these parameters are specially affected by the approximate rank deficiency (\tablename~\ref{tab:relpar}, compare columns 3 and 4)\footnote{In these simulations we did not solve for the parameter $GS_{\odot}$, as it was not originally included in our experiments. In this way, the comparison of the parameters' uncertainties in the table and the ones presented in previous papers is more straightforward}. 

The approximate symmetries that were identified to be responsible for the approximate rank deficiencies are approximate versions of the exact symmetries described in Sect.~\ref{subsec:adv_rk}, namely the rotational and the scaling symmetry. In all the simulations carried out so far, we adopted the common techniques used to cure rank deficiencies. In \citet{SCHETTINO-TOMMEI_2016}, for instance, the authors opted for descoping (four parameters were eliminated by the fit, namely the three coordinates of the EMB velocity and one coordinate of the EMB position), whereas, in \citet{SCHETTINO-etal_2016}, the authors introduced a set of apriori constraints, i.e., some mathematical relations among the solve-for parameters, aiming to inhibit the effect of the approximate symmetries. The presence of such constraints, though, does not seem to improve the uncertainty of the PPN parameters (\tablename~\ref{tab:relpar}, compare columns 3 and 5).  

The presence of an approximate rank deficiency of order 4 should reflect into the presence of 4 small (in some sense to be clarified) eigenvalues of the normal matrix. However, by performing a spectral analysis of such matrix, no evidence of small eigenvalues can be found (\figurename~\ref{fig:histfull}).

\begin{figure}[t]
\includegraphics[scale=0.6]{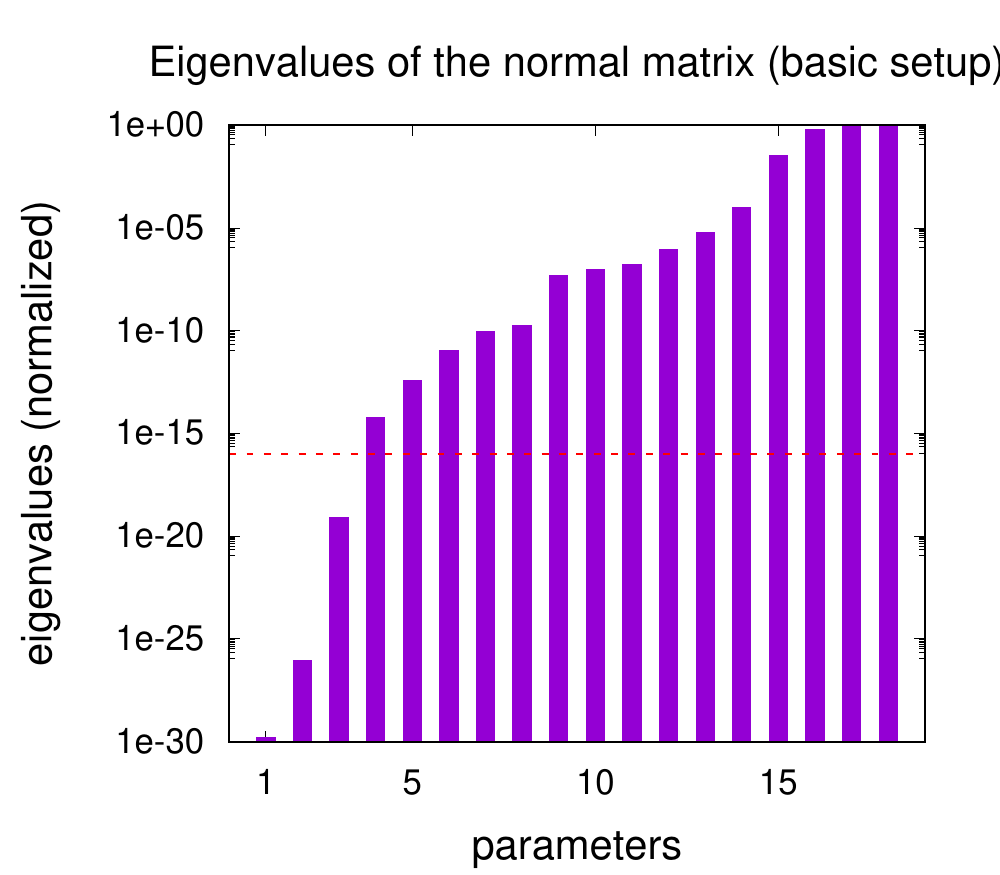}
\includegraphics[scale=0.6]{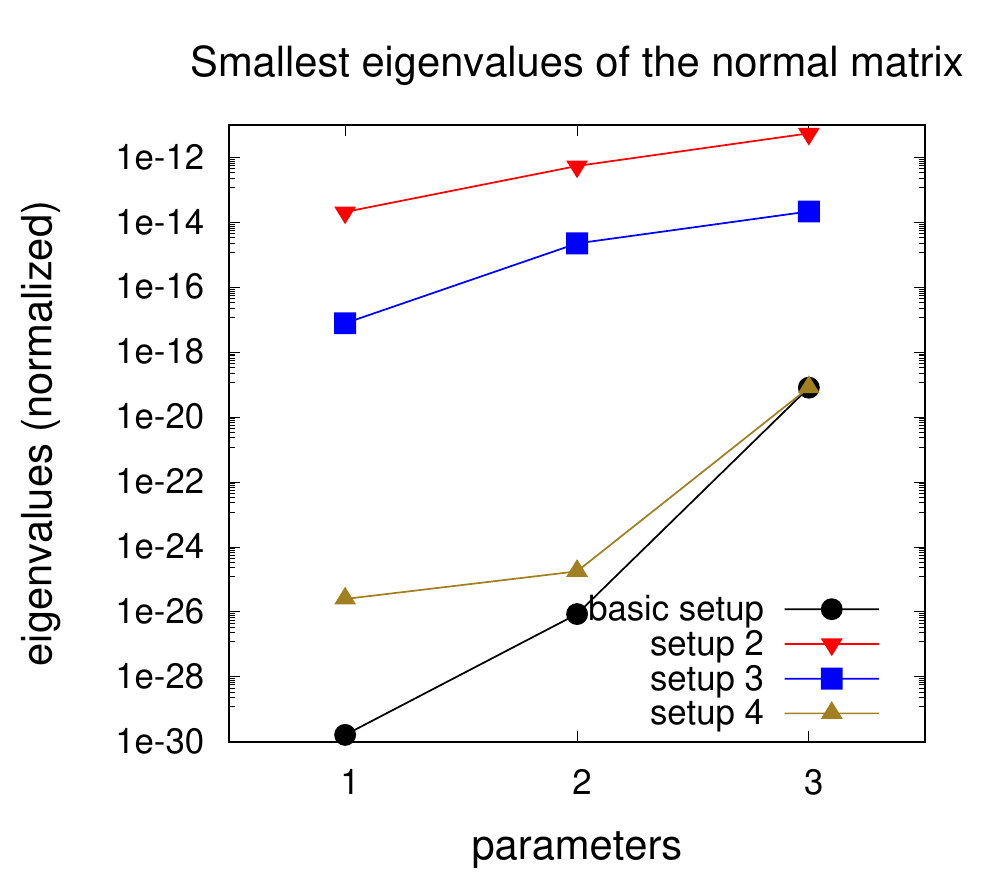}
\caption{On the left, ordered eigenvalues of the normal matrix in the case of the basic setup of our toy-model (see the text for the detailed description). On the right, comparison of the smallest three eigenvalues in the case we take into account also the rotation of the Earth (setup 2), the perturbation of the Moon (setup 3) or the perturbation of planets (setup 4).}
\label{fig:histcomp}
\end{figure}

Therefore, in order to investigate the issue of the presence of the approximate rank deficiencies in the MORE experiment in a more quantitative way, we started by simulating a simplified experiment, in which we are sure of the presence of exact symmetries. The set-up is the following: Mercury and the Earth (no Moon) as point masses orbiting the solar system barycenter, and a spacecraft orbiting Mercury in a keplerian orbit. Since the relativity experiment relies mainly on range data, in this basic setup we include only them in the fit. The solve-for parameters are the initial conditions of the Earth, Mercury and the spacecraft. In such configuration, we calculate the eigenvalues of the normal matrix and find that there are indeed three eigenvalues much smaller than $10^{-16}$, the machine precision (see \figurename~\ref{fig:histcomp}). From the spectral analysis there is no evidence of a fourth symmetry. As a matter of fact, the scaling symmetry should only be present if the observables are angles, which is not the case of BepiColombo, where we consider relative distances and velocities.  

In order to understand whether the order-3 symmetry breaks into an approximate one, we modified the basic setup described above by introducing three different perturbations, which in our software are computed through the call of external ephemerides and therefore are not affected by the action of the symmetry group. We introduce these effects one at the time, so as to evaluate their impact on the symmetry separately. The three effects that we selected are: Earth's rotation, which changes the position of the antenna; the presence of the Moon, which influences the orbit of the Earth-Moon barycenter; the gravitational attraction of the other planets, which influence the orbits of Mercury and the EMB. 

In \figurename~\ref{fig:histcomp}, we reported the three smallest eigenvalues of the normal matrices obtained with the different setups introduced above. The most significant impact on the magnitude of the eigenvalues is given by the Earth's rotation: the smallest eigenvalues in this case is about $10^{-13}$. This might be due to the fact that among the tested effects, the rotation of our planet is the one with the shortest period. On the contrary, the gravitational attraction of the other planets does not change much the eigenvalues: the largest of the three smallest eigenvalues is well below the $10^{-16}$ limit. 

In conclusion, we can say that the MORE experiment surely is not characterized by a scaling symmetry, neither exact nor approximate. Moreover, the effect of Earth's rotation seems to destroy completely the rotation symmetry, which does not even seem to be present in an approximate form. If there are no evident approximate rank deficiencies, it follows that there is no other way to improve the formal uncertainties of the PPN parameters than by changing the experiment, or by adding more observations. This option is currently being studied and was discussed in \citet{DEMARCHI-CASCIOLI_2020}.

\section{Conclusion}
\label{sec:concl}
In this paper, we presented the mathematical methods for performing radio-science experiments. First, we described the basic theory of OD, then we passed to more advanced tools necessary for dealing with interplanetary missions. The extraordinary accuracy of the data requires models increasingly sophisticated, both for the observations and the dynamical equations. All the elements described in the article have been implemented in the precise OD software Orbit14. The development of the code has taken several years and it is still going on, as the processing of real data always places new challenges and prompts to investigate new dynamical effects. Having full access to the source code, we can directly update the program according to the increasingly demanding requirements of the experiments. In this way, apart from managing to align the performances of Orbit14 with those of fully-tested and versatile OD software products, such as MONTE by JPL/NASA, we are able to carry out independent and innovative analyses.

In Sect.~\ref{sec:juno}, we showed that the code has succeeded in dealing with real data from a space mission. In particular, since 2016, the CMG of Pisa is processing the radio-science data of the NASA Juno mission using Orbit14. \citet{SERRA-etal_2019} presented a detailed analysis of OD experiments with the Doppler data from the first Juno gravity perijoves. In this article, we presented the results of OD experiments with the data from the first half of the mission, as previously done by \citet{DURANTE-etal_2020}. We found the same features for the gravitational and tidal field of Jupiter, showing further details of the Juno gravity experiment. We demonstrated that Orbit14 is able to provide an independent estimation of the dynamical parameters of Jupiter, working in parallel with other research groups which employ the NASA OD software MONTE \citep{IESS-etal_2018,PARISI-etal_2020}.

In Sect.~\ref{sec:bc}, we presented the ongoing study of the ESA/JAXA BepiColombo mission. The spacecraft will arrive at Mercury only at the end of 2025, so that several studies and simulations are being carried out in order to support the future interpretation of the radio-science data. As we have already described, the MORE experiment is twofold: gravimetry and rotation experiments focus on the investigation of the interior of the planet, while the relativity experiment provides a new test to study fundamental physics. Orbit14 has been developed to deal with both of them, as showed in \citet{CICALO-etal_2016}. Moreover, the software capitalized previous works devoted to the cure of the rank deficiencies arising in the mission geometric configuration and to the improvement of the classic multi-arc strategy \citep{MILANI-etal_2002,ALESSI-etal_2012}. Recently, the CMG of Pisa has further focused on the relativity experiment, investigating the testing of alternative theories of the gravitation, such as torsion theories \citep{SCHETTINO-etal_2020}.

For the next future, as Juno continues its mission around Jupiter, the processing of its radio-science data goes on. The bigger amount of observations allows to estimate a larger set of parameters of interest, but it implies also further difficulties, as the multi-arc fit of tens of perijoves reveals details of the dynamics that require careful investigations. BepiColombo, instead, is just at the beginning of its mission and many years separate the spacecraft from its insertion in orbit around Mercury. However, in March 2021, the mission executed its first superior-solar conjunction, providing the possibility to perform a first test of the General Relativity.

Although Orbit14 has been developed with the aim to support the BepiColombo and Juno missions, through the years it has acquired a large and general basis, making the software suited also for future space missions. \citet{LARI-MILANI_2019} showed that, with some efforts, it is possible to employ Orbit14 for missions with very different setups, such as the JUICE mission, which is going to perform multiple fly-bys of the Galilean satellites of Jupiter. Therefore, in the next years, we expect to further develop the code and use it in new challenging space missions.
\vspace{\baselineskip}

\noindent\small{\textbf{Acknowledgements} This research was funded in part by the Italian Space Agency (ASI) through agreement no. 2017-40-H.0. The Orbit14 software has been developed with the financial support of ASI. We thank the Radio Science Laboratories of the Università di Roma La Sapienza and Università di Bologna for many fruitful discussions and for their constant support to our work. We wish to dedicate this work to the memory of our mentor Prof. Andrea Milani Comparetti.}\\

\noindent\small{\textbf{Conflict of interest} The authors declare that they have no conflict of interest.}\\

\noindent\small{\textbf{Data availability} All Juno gravity data used in this article are publicly available through NASA-PDS repository at https://atmos.nmsu.edu/PDS/data/jnogrv\_1001/DATA. BepiColombo data were simulated starting from the SPICE kernels of the mission available through ESA-COSMOS repository at https://www.cosmos.esa.int/web/spice/bepicolombo.}

\bibliographystyle{apalike}
\bibliography{o14}

\end{document}